\documentclass{aa}

\usepackage{graphicx}
\usepackage[varg]{txfonts}

\usepackage{placeins}

\begin{document}

 
   \title{Exploring X-ray variability with unsupervised machine learning}
   \subtitle{I. Self-organizing maps applied to XMM-Newton data}

   \author{
        M. Kovačević \inst{\ref{ist1}}
          \and
        M. Pasquato \inst{\ref{ist4}, \ref{ist4.1}, \ref{ist4.2}, \ref{ist4.3}, \ref{ist1}}
          \and
        M. Marelli \inst{\ref{ist1}}
          \and
        A. De Luca \inst{\ref{ist1}, \ref{ist2}}
          \and
        R. Salvaterra \inst{\ref{ist1}}
          \and
        A. Belfiore \inst{\ref{ist1}}
          }

 \institute{
    Istituto di Astrofisica Spaziale e Fisica Cosmica (INAF IASF-MI), 20133 Milano, Italy\label{ist1}\\
              \email{milos.kovacevic@inaf.it; milosh.kovacevic@gmail.com}
         \and 
 Center for Astro, Particle and Planetary Physics (CAP$^3$), New York University Abu Dhabi \label{ist4}\\
             \email{mp5757@nyu.edu}
         \and
 Physics and Astronomy Department Galileo Galilei, University of Padova, Vicolo dell’Osservatorio 3, I–35122, Padova \label{ist4.1} 
         \and 
 Département de Physique, Université de Montréal, Montreal, Quebec H3T 1J4, Canada \label{ist4.2}
         \and 
 Istituto Nazionale di Fisica Nucleare - Sezione di Padova, Via Marzolo 8, I–35131 Padova, Italy \label{ist4.3}
         \and 
    INFN, Sezione di Pavia, via A. Bassi 6, I-27100 Pavia, Italy\label{ist2}             }

   \date{Received October 14, 2021 / 
        accepted January 26, 2022}

  \abstract
   {{\it XMM-Newton} provides unprecedented insight into the X-ray Universe, recording variability information for hundreds of thousands of sources. Manually searching for interesting patterns in light curves is impractical, requiring an automated data-mining approach for the characterization of sources.}
   {Straightforward fitting of temporal models to light curves is not a sure way to identify them, especially with noisy data. We used unsupervised machine learning to distill a large data set of light-curve parameters, revealing its clustering structure in preparation for anomaly detection and subsequent searches for specific source behaviors (e.g., flares, eclipses). }
   {Self-organizing maps (SOMs) achieve dimensionality reduction and clustering within a single framework. They are a type of artificial neural network trained to approximate the data with a two-dimensional grid of discrete interconnected units, which can later be visualized on the plane. We trained our SOM on temporal-only parameters computed from $\gtrapprox 10^{5}$ detections from the EXTraS (Exploring the X-ray Transient and variable Sky) catalog.}
   {The resulting map reveals that the $\approx 2500$ most variable sources are clustered based on temporal characteristics. We find distinctive regions of the SOM map associated with flares, eclipses, dips, linear light curves, and others. Each group contains sources that appear similar by eye. We single out a handful of interesting sources for further study.}
   {The condensed view of our dataset provided by SOMs allowed us to identify groups of similar sources, speeding up manual characterization by orders of magnitude. Our method also highlights problems with fitting simple temporal models to light curves and can be used to mitigate them to an extent. This will be crucial for fully exploiting the high data volume expected from upcoming X-ray surveys, and may also help with interpreting supervised classification models.}


   \maketitle


\section{Introduction}
\label{intro}

X-ray astronomy probes highly diverse phenomena related to the most extreme physical conditions observable in the Universe: very strong gravitational and/or electromagnetic fields, very high temperatures, and populations of particles moving close to the speed of light. Variability as a function of time is the rule in the X-rays, and studying the temporal properties of the sources is crucial to understanding their physics. The current generation of space-based X-ray observatories, by performing single-photon spectral imaging over a relatively large field of view, collect an enormous amount of information on hundreds of new serendipitous sources and their variability each day. 

The European Photon Imaging Camera (EPIC) on board the European Space Agency (ESA) X-ray Multi-Mirror Mission (XMM-Newton) spacecraft \citep{2001A&A...365L...1J}, consisting of two MOS\footnote{Metal Oxide Semi-conductor.} cameras \citep{2001A&A...365L..27T} and one pn detector \citep{2001A&A...365L..18S}, is the most powerful tool currently available with which to study the soft X-ray sky thanks to the unprecedented combination of a large field of view, high sensitivity to point sources, and good time resolution. More than 20 years since its launch, it is still fully operative. Based on its serendipitous data, a very rich catalog of X-ray sources has been produced, including more than half a million unique sources. The long time actively spent in orbit (exposure time of $\sim300$ million seconds up to now, with the prospect of further years of observations) guarantees unprecedented sky coverage for an X-ray telescope and the possibility of discovering relatively rare events.

All available temporal domain information were extracted for serendipitous {\it XMM-Newton} sources within the EU-FP7 EXTraS project \citep[Exploring the X-ray Transient and variable Sky;][]{2021A&A...650A.167D}. We characterized the aperiodic, short, and long-term variability (on timescales ranging from the EPIC time resolution\footnote{The pn detector has the time resolution of 73 ms.} to years) and searched for periodicity in more than 300\,000 unique sources; we also searched for fast transients in all observations. All EXTraS results are available in the EXTraS Public Archive. These include short-term and long-term light curves, power spectra, and a database of synthetic parameters (several hundred for each source, quantifying and describing all aspects of temporal variability). The potential of these results for science is very large for all classes of X-ray sources -- from the detection of a superflare from a nearby ultracool L dwarf star \citep{2020A&A...634L..13D}, to the observation of a supernova shock breakout in a distant galaxy at $z\approx0.1$ \citep{2020ApJ...898...37N}, to the discovery of pulsations in ultraluminous X-ray sources \citep{2017Sci...355..817I, 2017MNRAS.466L..48I}. 

The EXTraS project is an example of astronomy entering into the big data era. There are at least two ways in which a data set can be considered "big": because it contains many objects, such as stars or galaxies, and because for each of these objects a large number of attributes has been measured. Imagined as a table, the first case corresponds to a large number of rows, and the second to a large number of columns, resulting in a high-dimensional data set. While, traditionally, authors have had to deal with the problem of having too little data, the era of big data poses a set of new, complementary problems. Condensing a data set by reducing its size becomes useful and even necessary \citep[see e.g.,][]{bien2011prototype}. Many unsupervised methods in machine learning focus on this exact task: "clustering" attempts to reduce the number of rows, extracting or synthesizing a limited number of representative instances; variable selection and "dimensionality reduction" on the other hand attempt to reduce the number of columns by either selecting few relevant variables, or by combining several variables into new ones.

In this paper we make use of a technique that accomplishes both dimensionality reduction and clustering at the same time: self-organizing maps \citep[SOM;][]{kohonen1982self, Kohonen2001}. This technique identifies groups of sources with shared characteristics, mapping them out onto a plane. This allows us to optimize visual inspection of the sources, revealing groups that share astrophysically relevant behavior (e.g., flares, eclipses) despite the fact that the method is agnostic with respect to the underlying physics.

While this approach is very well suited to our data, a broad variety of machine learning techniques, both unsupervised and supervised, is being increasingly applied to astronomy. The former are concerned with extracting patterns from a data set without direct guidance in the form of labeled data, while the latter focus on learning a function from labeled examples to carry out classification or regression. Examples of the former are anomaly detection \citep{2006MNRAS.369..677P, 2017MNRAS.465.4530B, 2020MNRAS.499..524G}, clustering \citep[e.g.,][]{2019MNRAS.490.3392P}, dimensionality reduction \citep[e.g.,][]{2018MNRAS.476.2117R}, and even integrated approaches including interactive visualization \citep[][]{2021A&C....3400437R}. While we do not discuss supervised methods  in the following (nor even unsupervised methods except for SOM), we point the interested reader to two relevant reviews: \cite{ball2010data} and the more recent \cite{2019arXiv190407248B}.

The paper is organized as follows. In Sect.~\ref{SOM}, we give a detailed explanation of our unsupervised learning approach, in Sect.~\ref{data} we described our dataset, in Sect.~\ref{results} we present our results, and in Sect.~\ref{conclusions}  we draw conclusions.


\section{Self-organizing maps}
\label{SOM}

\subsection{General information}
\label{som_info}

A SOM is a type of artificial neural network (ANN), but despite this classification, SOMs work quite differently from typical ANNs such as feed-forward neural networks and related architectures\footnote{For example, convolutional neural networks, etc.}. Also, unlike most ANNs, SOMs are designed for unsupervised learning tasks, performing dimensionality reduction and clustering for data visualization.

Self-organizing maps have already found wide application in astronomy, especially when dealing with large multidimensional data sets. They have been applied: to light curves of variable stars \citep{2004MNRAS.353..369B, 2016MNRAS.456.2260A}; as an aid in the context of photometric redshift estimation \citep{2012MNRAS.419.2633G, 2015ApJ...813...53M}; to cluster gamma-ray bursts \citep{2002ApJ...566..202R}; for morphological classification of galaxies \citep{1997ApJS..111..357N}; to find star clusters or otherwise coherent structures in Gaia data \citep{2018ApJ...863...26Y, 2020ApJ...891...39Y, 2020ApJ...900L...4P}; to find anomalous data in SDSS spectra \citep{2013A&A...559A...7F, 2014A&A...568A.114M}; to find variable active galactic nuclei \cite[][]{2019ApJ...881L...9F}, and so on.\\

\begin{figure}[ht]
    \centering
    \includegraphics[width=0.95\columnwidth]{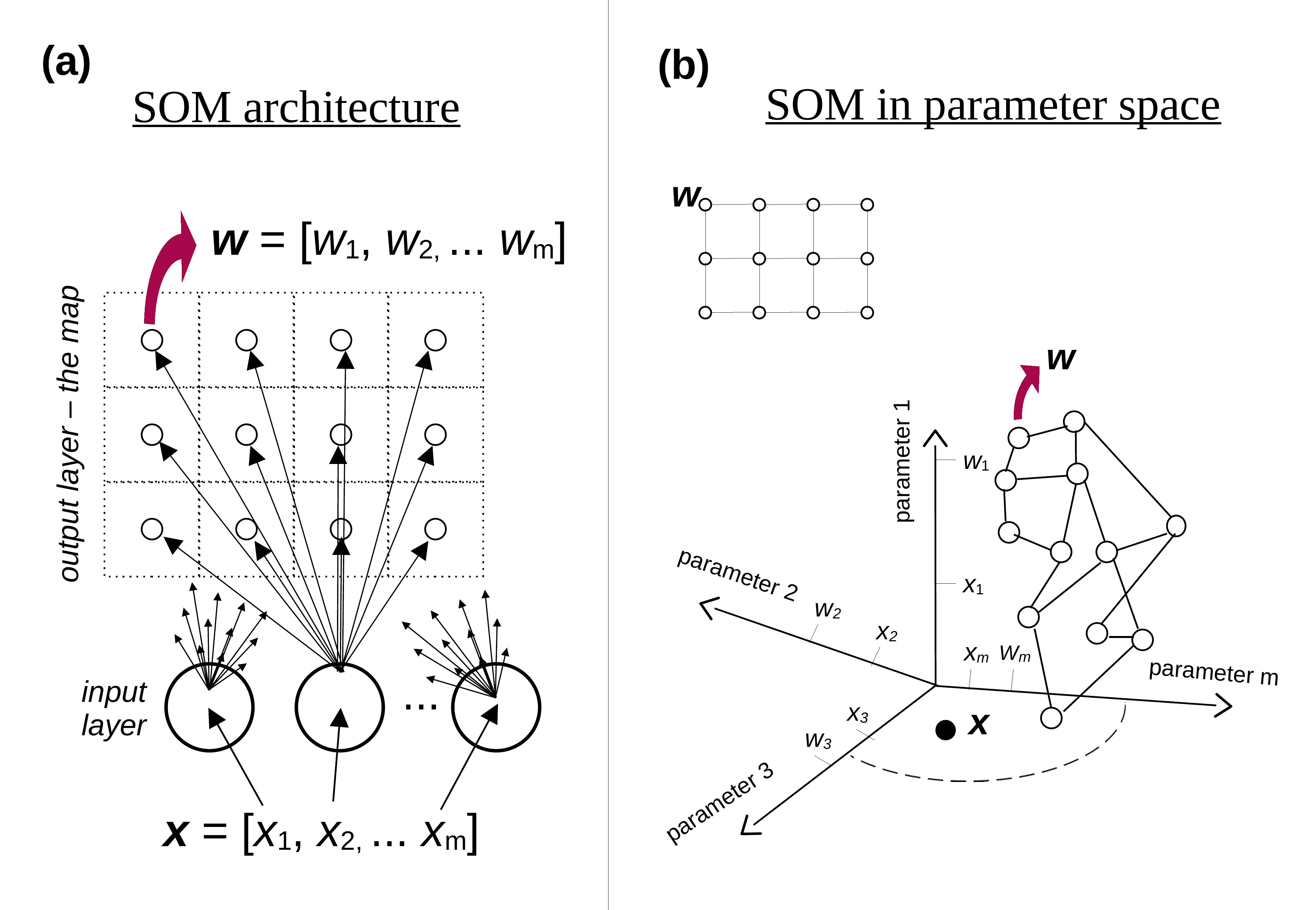}
    \caption{
    SOM schematics. \protect\\
    \textit{Panel a}. SOM architecture. The bottom circles represent input neurons while the upper smaller circles represent the output neurons which build up the flat 2D map. The input neurons receive the values of the object parameters and are connected to all the output neurons as indicated through arrows. The dashed squares centered on the output neurons are the pixels used to visualize the map. For the sake of clarity, only one input neuron shows arrows going all the way to each output neuron. Also weights $\vec{w}$ of only one output neuron are explicitly shown. \protect\\
    \textit{Panel b}. SOM in $m$-dimensional parameter space. The position of the output neurons on the flat 2D map grid is visualized with lines connecting them. The same map is shown below, immersed in the input parameter space, where each axis represents one parameter. The coordinates of one neuron $\vec{w}$ and one object $\vec{x}$ are explicitly shown on the axes. 
    }
    \label{fig:som}
\end{figure}

The SOM architecture is simple, consisting of an input layer and an output layer (Fig.~\ref{fig:som}; panel~a). The input layer consists of $m$ neurons, where $m$ is the number of input parameters (one neuron per parameter). Each neuron in the input layer is connected to all the neurons in the output layer. The output layer is typically a 1D, 2D, or 3D\footnote{For visualization purposes, the output layer has a maximum of three dimensions. However, as a dimension reduction algorithm, the SOM output layer can have any number of $\leq m$ dimensions.} network of neurons connected to each other in the form of a grid.

The output layer is the place where visualization, dimensionality reduction, clustering, and so on is observed and presents the actual map. 
Typically, a flat 2D map is used and the shape is usually rectangular (four edges)\footnote{It can also be cylindrical (two edges), or closed, as in a sphere, an ellipsoid, or a torus surface.}. The output of these neurons indicates the number of objects placed on them. It is visualized as a map consisting of pixels where each pixel corresponds to a neuron. The shape of a pixel is typically square or hexagonal. 

Each neuron in the output layer (the flat 2D map) has a unique set of $m$ weights associated to it: $\vec{w} = [w_1, w_2, ... w_m]$ (Fig.~\ref{fig:som}; panel~a). The number of weights $m$ is the same as the number of input parameters describing each object $\vec{x} = [x_1, x_2, ... x_m]$. When a certain object with parameters $\vec{x}$ is presented to the input layer, that object is placed on the neuron whose weights $\vec{w}$ are most similar to the input parameters $\vec{x}$. The most commonly used metric for this purpose is the Euclidean distance:
\begin{equation}\label{eq:d1}
d = ||\vec{w} - \vec{x}|| = \sqrt{ (w_1 - x_1)^2 + (w_2 - x_2)^2 + ...~(w_m - x_m)^2 }~.
\end{equation}
The object is placed on the neuron with the smallest $d$. This neuron is commonly referred to as the "best matching unit" (BMU). When all $n$ objects in the sample are presented to the SOM, they are distributed across the map depending on the weights of each neuron. 

Another way to think about assigning objects to neurons is to imagine an Euclidean $m$-dimensional parameter space (Fig.~\ref{fig:som}; panel~b). The objects parameters $\vec{x}$ are then coordinates in this space and all objects populate this space. The neuron map coordinates in this space are their weights $\vec{w}$. The map grid, that is, positions of neighboring and other neurons on the flat 2D map, can be seen as the lines connecting them. This map is a curved discrete 2D surface embedded in the $m$-dimensional parameter space. Each object is assigned to its closest neuron according to the Euclidean distance metric $d$.

Assigning objects to neurons in the map represents a dimensionality reduction. Each object with $m$ parameters associated to it has only two discrete parameters on the 2D map (the map grid coordinates of its BMU).

SOMs are designed to detect patterns, clusters, and so on of objects based on their parameters and to preserve the topology of their $m$-dimensional distribution when placing them on the map. Objects that are similar to each other (nearby in parameter space) should be close to each other on the map. If there are distinct groups of objects, it should show up on the map as 2D groups. If the map weights are random, the objects will be randomly distributed on the map, and so the map needs to be trained, that is, its weights adjusted according to the parameters of the objects. In this sense, the SOM algorithm is similar to other ANN algorithms which also need to be trained. However, objects used for training the SOM are not labeled. The objects do not necessarily need to be divided into training and testing (and validation) samples, and there is no loss or cost function that needs to be minimized until it converges to a global minimum. It is possible to define a certain cost function for a SOM and monitor its reduction as training progresses, but the minimization of that function is not behind the training algorithm.\\

The way the SOM algorithm works is by introducing objects, finding their BMU, adjusting the weights of the BMU to more closely  match the parameter values of the object, then doing the same to the weights of the surrounding neurons, but to a lesser degree the further they are on the flat 2D map grid. The last part is essential for the self-organizing property of the map and enables similar objects to be placed on nearby pixels on the trained map. It also means that the positions of neurons on the flat 2D map grid with respect to each other is important when training the map. ~Another important factor is that with each iteration (presentation of object(s) to the algorithm) the weight adjustment and the radius around the BMU are reduced. This allows the map to settle to the final position after a sufficient number of iterations. ~Finally, the weight adjustment depends linearly on the difference between weights and respective object parameter values, ensuring that the update of weights towards parameter values is larger when the difference between them is larger. The dependence of weight adjustments on the (flat 2D map) distance from the BMU, $l$, is described by a "neighboring function", $h$. Typically $h$ is a 2D symmetric Gaussian function centered on the BMU:
\begin{equation}\label{eq:h1}
h = \exp(-~\dfrac{l^2}{2\sigma_i^2})~.
\end{equation}
The $\sigma_i$ factor controls the width (standard deviation) of the neighboring function $h$ which reduces with each iteration $i$, typically in a exponential manner.
The formula (in vector form) for updating the weight $\vec{w}$ of a neuron at a distance $l$ from the BMU is:
\begin{equation}\label{eq:w1}
\vec{w}(i+1) = \vec{w}(i) + \alpha(i) \times h(l,\sigma(i)) \times [\vec{x} - \vec{w}(i)]~.
\end{equation}
At each iteration $i,$ the object $\vec{x}$ is different\footnote{Therefore, the distance from the BMU on the flat 2D map $l$ also depends indirectly on the iteration $i$ because a new object \vec{x} mainly corresponds to a different BMU.}
until all $n$ objects from the sample are passed. This completes one epoch of training. The total number of iterations $i_{max}$ is then $i_{max} = n \times n_{ep}$ where $n_{ep}$ is the number of epochs. The term $\alpha$(i) is chosen such that, at $i=0,$ it starts from a certain maximal value and decreases to a certain minimal value at $i=i_{max}$ which is usually significantly smaller than the starting value. The term $\sigma$(i) typically starts from a value similar to the size of the map at $i=0$ and decreases to encompass just one neuron (the BMU) at $i=i_{max}$.

Some conclusions can be drawn from the above algorithm, which is referred to as the "online algorithm."~
As mentioned before, objects are not labeled and there is no cost function minimization behind this algorithm. The number of iterations is predetermined and does not depend on the cost function converging to a minimum. The way $\alpha(i)$ and $\sigma(i)$ are defined ensures the convergence of the map. However, it is important that the number of epochs is large enough for the map to converge smoothly to its optimal stage.~
Even starting from the same initial weight values, the final map will be different if the order of introducing objects in the sample is changed. In this case the final map should still show the same groups and patterns, but these will be located in different places on the map.~
Each iteration can only be performed after the previous one is completed. Therefore, the algorithm is one large loop and the process cannot be parallelized and remains relatively slow.

A similar form of algorithm also exists, called the "batch algorithm," which processes all objects in the sample at the same time for each epoch. The formula that regulates how weights are updated is:
\begin{equation}\label{eq:w2}
\vec{w}(i) = \dfrac{ \sum_{j=1}^{j=n} h(l_j,\sigma(i)) \times \vec{x}_j }
{ \sum_{j=1}^{j=n} h(l_j,\sigma(i)) }~.
\end{equation}
In this case, each iteration $i$ represents one epoch, meaning that the total number of iterations is the number of epochs $i_{max} = n_{ep}$. The summation is over all $n$ objects in the sample. The factor $\sigma(i)$ changes only between epochs and there is no term $\alpha(i)$ as in the online algorithm. Within the summation, the term $l_j$ depends on the object $\vec{x}_j$ and its BMU.

Again, some conclusions can be drawn.~
As in the online algorithm, there are no labeled objects and no cost function minimization; the map converges to the final position on its own, and simply needs enough epochs to converge to an optimal state.~
If the initial weight values are the same, the final map will be the same regardless of the order of the objects in the summation.~
The summation part can be parallelized and the only serial loop is over the epochs. This can make the batch algorithm faster and saves time, which can make a significant difference if there are many objects $n$ in the sample.

Equation (\ref{eq:w2}) can be made more algorithmically concise by grouping objects with the same BMU together and summing over each pixel:
\begin{equation}\label{eq:w2b}
\vec{w}(i) = \dfrac{ \sum_{k=1}^{k=n_{pix}} n_k^{bmu} \times h(l_k,\sigma(i)) \times \vec{\overline{x}}_k }
{ \sum_{k=1}^{k=n_{pix}} n_k^{bmu} \times h(l_k,\sigma(i)) }~.
\end{equation}
Here the $n_{pix}$ is the number of pixels, $n_k^{bmu}$ is the number of objects whose BMU is pixel $k$. The term $l_k$  remains the same for all objects with the same BMU. The term $\vec{\overline{x}}_k$ is an average vector value of all objects with the same BMU. This can further speed up the algorithm.

\subsection{The algorithm}
\label{som_alg}

In this work, we used {\it SOMPY}\footnote{\url{https://github.com/sevamoo/SOMPY}} \citep{moosavi2014sompy} for the SOM implementation. It is written in Python\footnote{\url{https://www.python.org}}, uses batch training (Eqs.~\ref{eq:w2} and \ref{eq:w2b}), and is relatively fast. The algorithm characteristics and options are as follows, along with the settings chosen for this paper.~
It uses a flat 2D rectangular map with either square or hexagonal pixels: we chose square pixels for simplicity.~
We fixed the number of output pixels to the default value of $5 \times \sqrt{n}$ ($n$ is the number of objects). This is a common rule of thumb regarding the map size.~
The map proportions were set to the default value, which was obtained from the proportional length of the two largest PCA\footnote{Principal component analysis.} vectors for the data set, which is another common rule of thumb for the map.~
The neighborhood function $h$ can either be Gaussian (Eq.~\ref{eq:h1}) or "bubble"\footnote{A radial 2D function with a constant value that drops to zero at a given radius.}. Gaussian function was chosen because it is typically used as a neighborhood function. The algorithm splits training in two parts: "rough" and "fine," each with its own number of iterations. This is related to the value of the average width $\sigma_i$ (Eq.~\ref{eq:h1}). During rough training, it starts from a value somewhat smaller than the map length and ends up with a value several times smaller. During fine training, it starts from the previous value and ends at a value close to $\simeq 1$, which is the distance between a BMU and its neighboring pixels\footnote{Up, down, left, and right, not the four diagonal.}. In both cases, $\sigma_i$ decreases linearly with each batch iteration $i$. By dividing the training phase into two parts with given starting and ending values for $\sigma_i$, it approximates exponential decay.~
Weight initialization can be random or defined by PCA. The second case initializes weights in such a way that the map forms a grid on a plane defined by the two largest PCA components in parameter space, and is centered on the data. This method was chosen because it gives a good starting position for the map even if the data are not intrinsically two-dimensional and linear.~
There are various options for normalizing the data, but a custom normalization was used, which is explained in the following section.
The number of training epochs for both rough and fine training were chosen such that the final map does not change significantly and that the average "quantization error" does not change by more than 1 per cent when doubling the number of epochs. The quantization error is the difference between parameter values of an object and the weights of its BMU defined as $d^2$ (Eq.~\ref{eq:d1}). The average quantization error is the average $d^2$ over all $n$ objects.


\section{Data selection}
\label{data}

Among the several results released by the EXTraS collaboration\footnote{\url{http://www.extras-fp7.eu/index.php/archive}}, we explored the catalog reporting the short-term aperiodic variability analysis. For each detection, several short-term (within the time-span of one orbital period ~$\la$ 160 ks) light curves are extracted and statistical parameters computed, where a detection is defined as an observation of a unique source within a unique {\it XMM-Newton} observation period\footnote{Period during which the telescope spends pointing in one direction.} with a unique camera\footnote{There are three cameras  in total: pn, MOS1, MOS2.} and within a unique exposure time during the observation period\footnote{There can be several exposure times for a single camera during one XMM-Newton observation period.}. There are four types of light-curve binning, six temporal models fitted to the light curves, and four energy bands. All of these combinations coupled with various other parameters extracted from light curves resulted in several hundred parameters for each detection.

The short-term variability EXTraS catalog comprises 872\,075 detections, each described through 754 parameters\footnote{See the help pages of the EXTraS short-term variability archive for a complete list and description of the parameters.}. Parameters were chosen such that they: were derived from light curves with one set of time-bin definitions; only contain variability (and not spectral) information; and do not have many "null" values. We excluded the count rate or any proxy for the count rate, and other parameters\footnote{Information related to identification, duration of observation, errors, redundant instrumental and statistical information, etc.}.

Starting from all of the 754 parameters, the selection criteria reduced their number in the following way.~
We only accepted parameters that were derived from light curves with uniform time bins of 500 s (down to 147 parameters).~
We selected light curves encompassing the full energy range, not any of the three subranges (down to 84 parameters).~
All the parameters related to the "exponential decay," "flare", and "eclipse" models were excluded because they contain many null values (down to 53 parameters). The parameter "relative excess variance" and its error were excluded for the same reason (down to 51 parameters).~
The parameter "average count rate" and its proxies\footnote{Median count rate and the first coefficients in the "constant," "linear", and "quadratic" models.} were excluded (down to 47).~
Finally, excluding other parameters related to identification and so on leaves $m = 31$ parameters.~
The final selection of these parameters and their description is presented in Table~\ref{tab:params} (Appendix~\ref{app:data}).

The 872\,075 detections in the catalog were filtered through flags and quality checks, requiring at least 20 time bins, non-negative count rate, and non-null values for all the 31 parameters. By combining all these constraints, we are left with $n = 128\,925$ detections.

The astrophysical type is unknown for a large number of {\it XMM-Newton} sources.
Our $n = 128\,925$ detections correspond to about $43,000$ unique sources, of which approximately $6000$ reside in the Galactic plane $|b| \le 2^{\circ}$ ($\sim 11\,000$ within $|b| \le 10^{\circ}$). Given that most stars are within the Galactic plane while AGNs are above it, this can give an approximate idea of the composition of our sources. For a more quantitative description, a recent classification of {\it XMM-Newton} sources using a supervised method \citep{2021arXiv211101489T} found that about $80\%$ of the sources are AGNs, $\sim 20\%$ are stars, and few per cent are X-ray binaries and cataclysmic variables. These proportions should be similar with our sources.

The normalization of the $m = 31$ parameters and their mutual correlation is nontrivial and is explained in detail in Appendix~\ref{app:data}. With the data set of $n = 128\,925$ detections (samples) and adopting the configuration options explained in Sect.~\ref{som_alg}, the SOM algorithm settings in this case are: a map size of $45 \times 40$; 80 training epochs for both rough and fine tune training; $\sigma_i$ (Eq.~\ref{eq:h1}) decreasing linearly from 6 to 1.5 during rough training and from 1.5 to 1 during fine training. With the batch implementation of the algorithm it took only about 5-10 minutes to train the algorithm on $n = 128\,925$ detections (samples) with $m = 31$ parameters (features) over 160 epochs (iterations) on an average CPU (one CPU at 2.6-3.5 GHz with four cores).


\section{Results}
\label{results}

\subsection{SOM applied to the EXTraS data}
\label{res:som_extras}

As explained in detail in Sect.~\ref{SOM}, SOM performs dimensionality reduction starting from {\it n} objects described by {\it m} parameters resulting in a 2D map (BMU map) populated by {\it n} objects. At the same time, it performs clustering, such that objects which are similar end up close to each other on the BMU map forming a group.

\begin{figure}[htb]
    \centering
    \includegraphics[width=0.95\columnwidth]{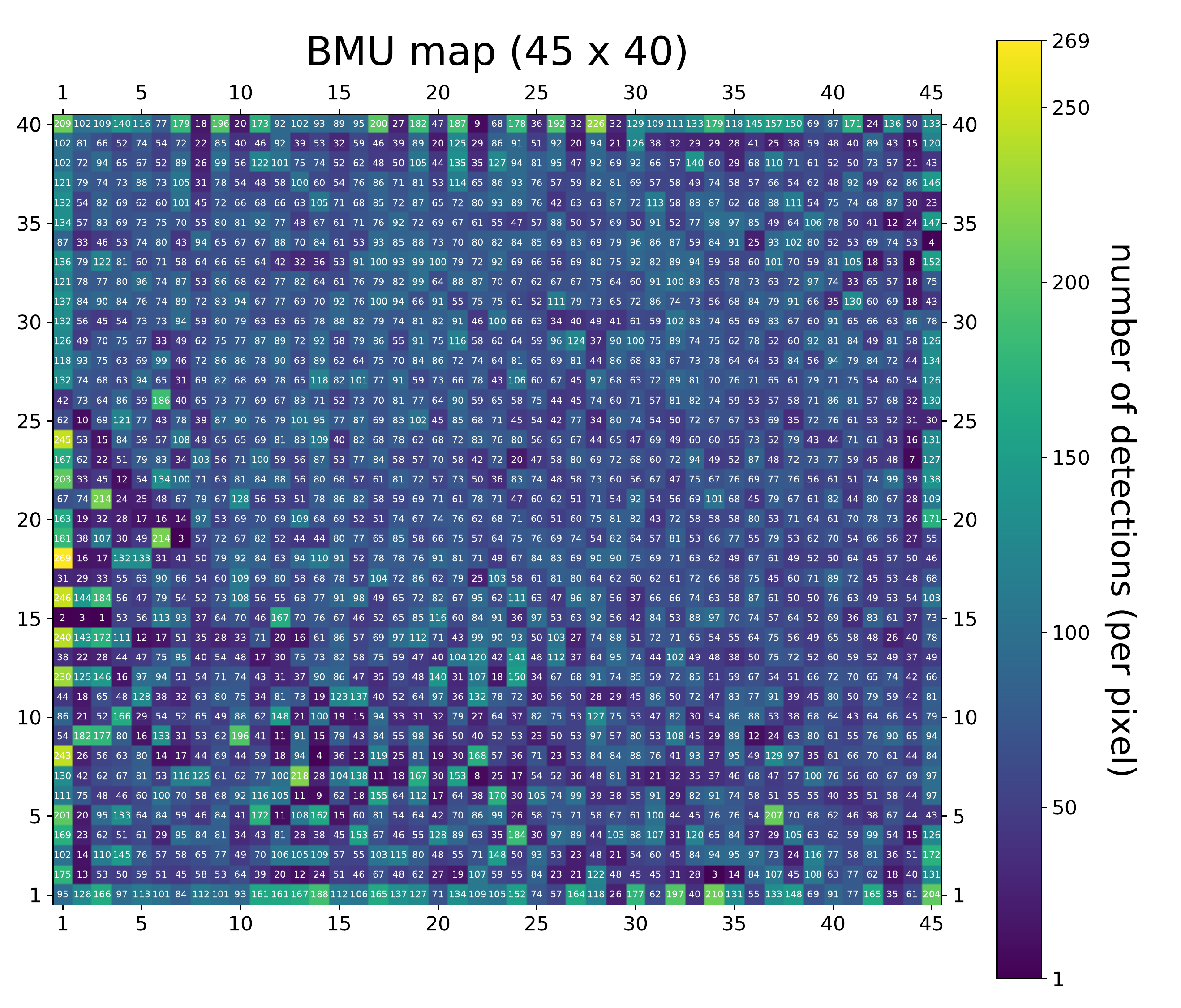}
    \caption{
    BMU map. This BMU map corresponds to all $n = 128\,925$ detections. The color bar measures the number of detections placed on each pixel. This value is also indicated as a number on top of each pixel. The coordinates of the pixels within the grid start from the lower-left corner and are indicated on all sides of the map.
    }
    \label{fig:bmu_tot}
\end{figure}

As reported in Sect.~\ref{data}, we applied SOM on $n = 128\,925$ {\it XMM-Newton} detections described by $m = 31$ variability parameters. The resulting BMU map is shown in Fig.~\ref{fig:bmu_tot}. The numbering of pixels starting from the bottom left was introduced for guidance.
The center of the map is mainly uniform while the lower-left part, which has a triangular shape, is highly fractured and seems to form a separate part. This suggests that the majority of detections in the broad map center form a single group in the normalized $m = 31$ dimensional parameter space. The lower-left part of the map suggests that detections placed here form many small groups. 

\begin{figure}[htb]
    \centering
    \includegraphics[width=0.95\columnwidth]{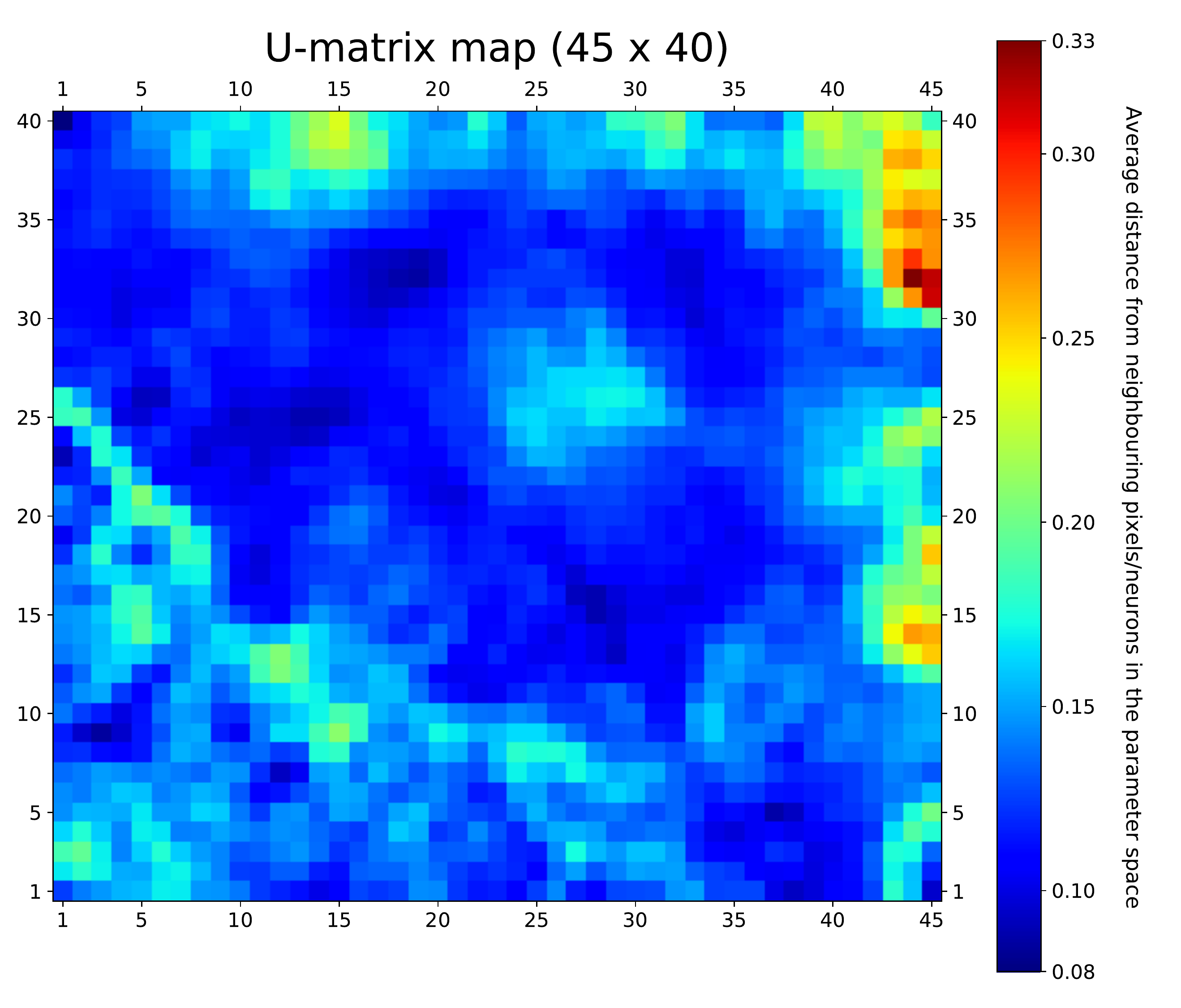}
    \caption{
    U-matrix map. The color bar indicates the average distance of each pixel-neuron to its neighboring pixels-neurons in the normalized parameter space. Further details are in the text.
    }
    \label{fig:umatrix}
\end{figure}

In Fig.~\ref{fig:umatrix} the U-matrix plot is shown, which allows us to identify the clustering structure of our data by displaying the distance of each neuron from its four nearest neighbors in the normalized parameter space. Groups of similar (nearby) neurons representing points in our normalized parameter space, whose dimension is $m = 31$, can thus be visualized as regions of nearby neurons in the plane. These appear as contiguous dark blue structures in Fig.~\ref{fig:umatrix}. Lighter colors, from sky blue to red, correspond instead to regions of lower density that divide groups (e.g., groups in the lower-left triangle; their division is seen as lines on the U-matrix map) or lie at the edges (e.g., groups on the U-matrix map at the up and right edges). The second case represents groups of outliers, that is, points that for whatever reason are different from the typical object in our data set. Clearly, objects that are systematically different may represent astrophysically interesting sources worthy of further study.

\begin{figure*}[htb]
    \centering
    \includegraphics[width=0.95\textwidth]{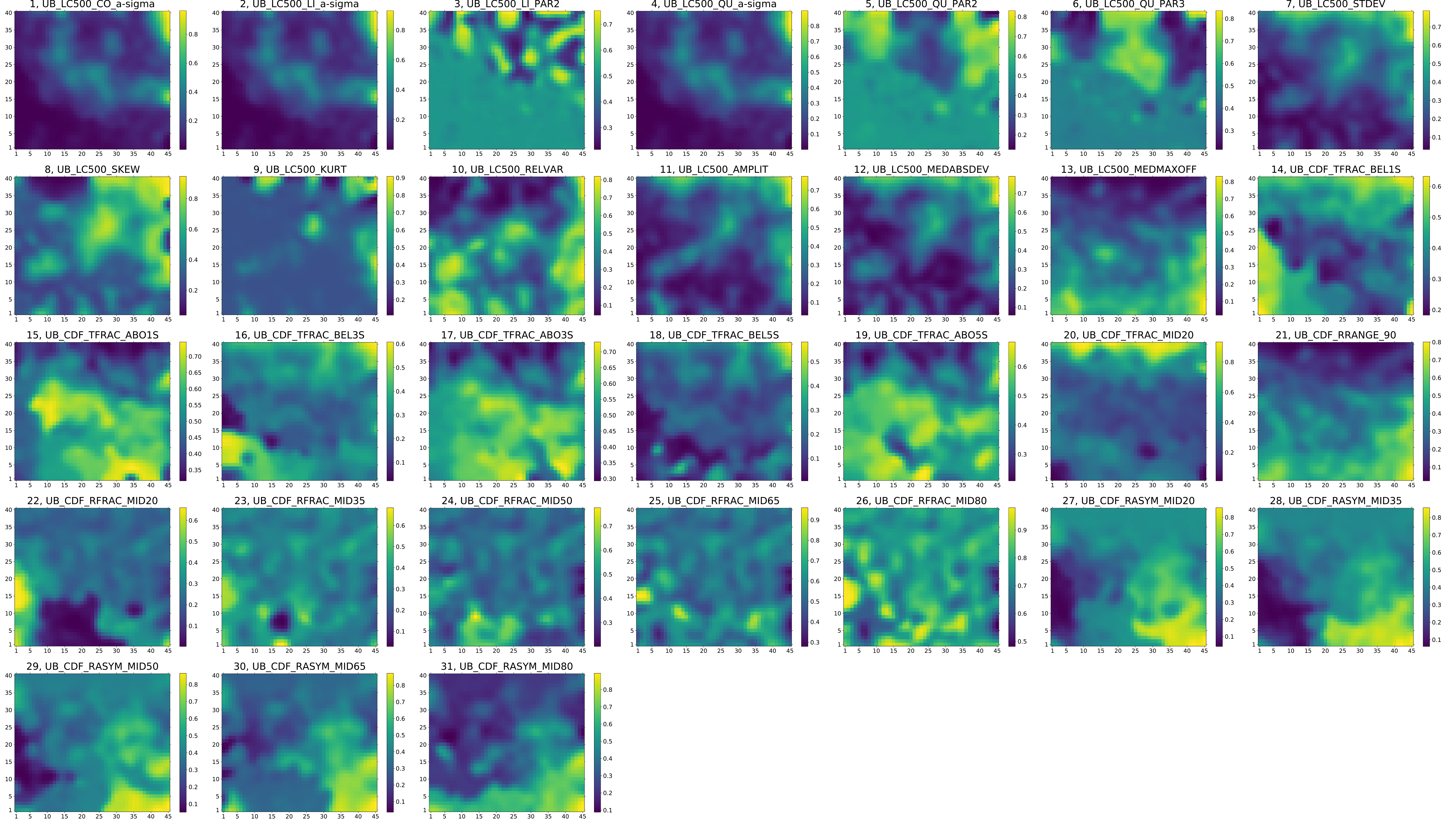}
    \caption{
    SOM weights. SOM weights for each parameter are presented as maps of the same dimension as the main BMU map. The color bar represents the value of the weight for each pixel. The numbering of the parameters is the same as in Table~\ref{tab:params}.
    }
    \label{fig:som_weights}
\end{figure*}

Figure~\ref{fig:som_weights} reveals how each of the $m = 31$ parameters map out onto this plane. Technically, these are the weights of the SOM neurons being shown over a grid of 1800 ($40 \times 45$) neurons. Even though there are $45 \times 40 = 1800$ neurons  in total and each of them has 31 weights for each parameter (corresponding to the relevant $m$ coordinates in parameter space), Fig.~\ref{fig:som_weights} shows that they can be easily visualized as a $45 \times 40$ map for each parameter (i.e., each coordinate in the parameter space).

These maps can be visually compared to Fig.~\ref{fig:umatrix} to reveal the characteristics associated to each data subgroup; for example it can be readily checked that the upper-right corner of the map (corresponding mostly to outliers according to Fig.~\ref{fig:umatrix}) has distinct variability properties. The combination of Fig.~\ref{fig:umatrix} and Fig.~\ref{fig:som_weights} thus acts as a look-up table guiding direct visual inspection of the sources.

\subsection{Analysis of variable sources}
\label{res:var_sources}

\begin{figure}[ht]
    \centering
    \includegraphics[width=0.95\columnwidth]{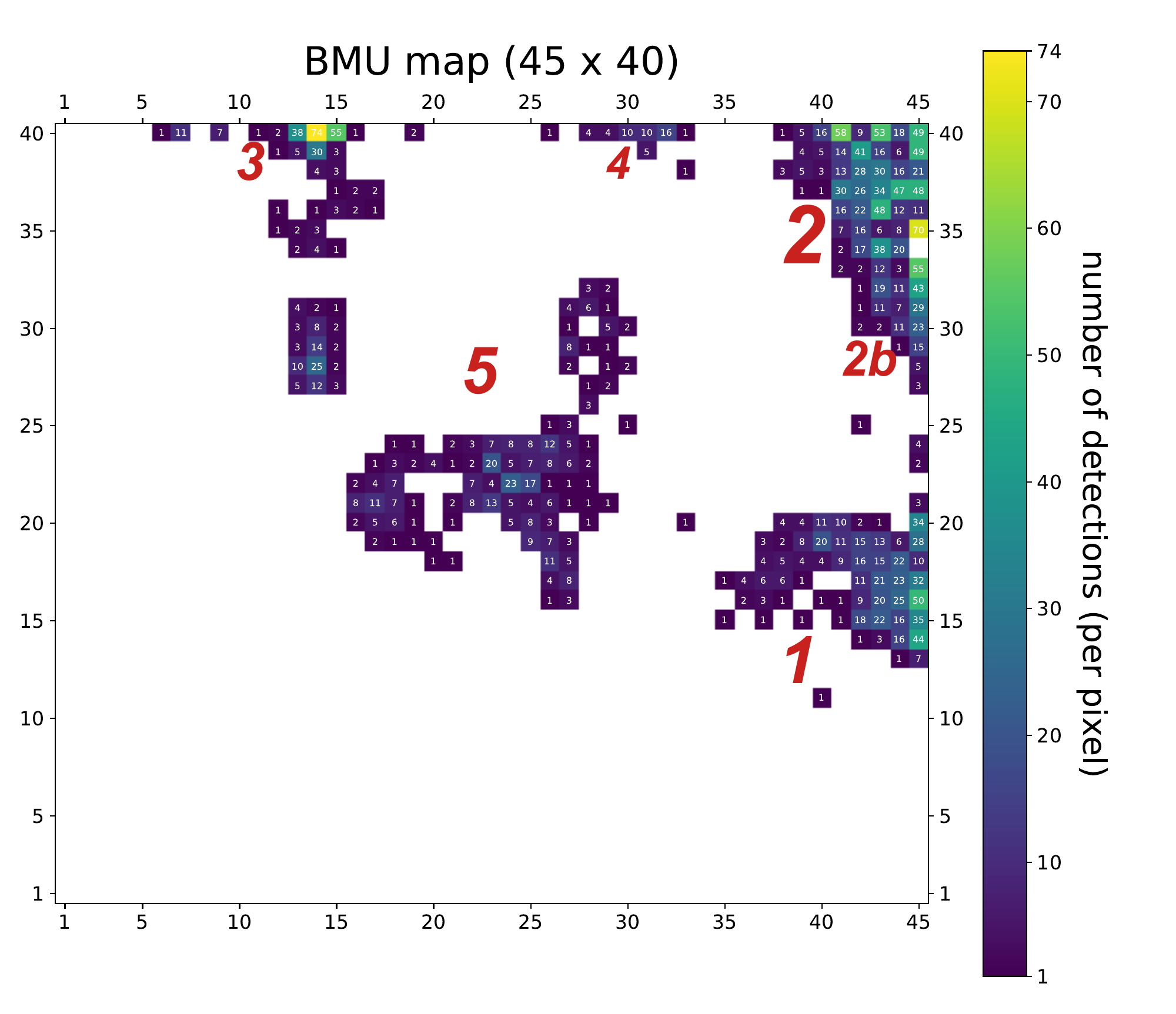}
    \caption{
    BMU map. Same map as in Fig.~\ref{fig:bmu_tot} but showing only the most variable $n_{var} = 2654$ pn detections. White pixels correspond to zero detections. Red numbers correspond to numbering of "blobs" (more in the text) and their size is illustrative of the number of detections in each one.
    }
    \label{fig:bmu_var}
\end{figure}

In order to examine potentially interesting detections more closely, we focused on variable sources. Variability was defined such that the fit of the 500 s time bin light curve with a constant model is unacceptable ($> 5\sigma$). Also, we only took the most sensitive pn camera light curves into account so as to ensure a sample with all the unique detections (unique source within unique time frame). The number of detections fulfilling these requirements is $n_{var} = 2654$. Their placement on the main BMU map (which was trained on all $n = 128\,925$ detections) is shown in Fig.~\ref{fig:bmu_var}.

It is apparent that variable sources form distinct groups, "blobs," separated from each other. Most groups take the same form as the core where the majority of detections are found, with the number of detections decreasing to zero as the distance from the core increases. Also, some groups show some substructure. In Fig.~\ref{fig:umatrix}, pixels with the highest value mostly correspond to pixels containing groups of variable detections. The high value in the U-matrix map means that these neurons are far away from the neighboring neurons in the parameter space.

\begin{figure*}[ht]
    \centering
    \includegraphics[width=0.95\textwidth]{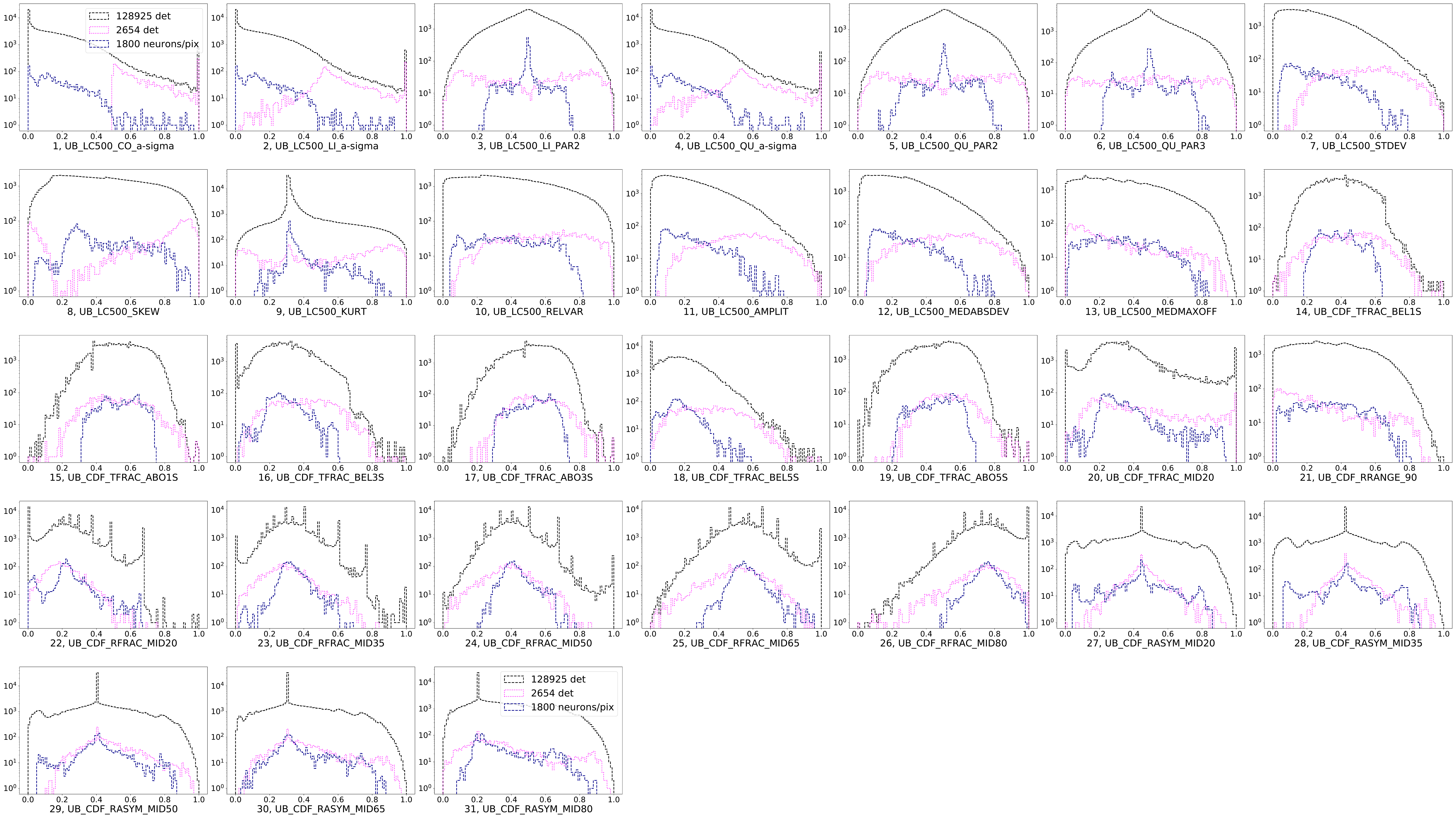}
    \caption{
    Distributions of values for each of the $m = 31$ parameters. Comparison between the distribution of the whole dataset (black) and that of the SOM neurons (blue) is shown. We additionally visualize the distribution of variable sources (pink). The numbering of the parameters follows Table~\ref{tab:params}.
    }
    \label{fig:som-units_x31}
\end{figure*}

Figure~\ref{fig:som-units_x31} shows that 1800 SOM neurons follow the distribution of all the $n = 128\,925$ detections rather well\footnote{The y-axis is in logarithmic scale.} for each of the $m = 31$ parameters. The distribution of the variable $n_{var} = 2654$ detections is more "stretched" towards the edges than that of all detections\footnote{This might be intuitively expected for the most variable detections.}. For most parameters at the very edges (near zero and one), the two distributions practically overlap. This means that variable detections tend to lie towards the edges of the parameter space. This can be interpreted as variable detections belonging to several quasi-outlier groups; this interpretation is confirmed by looking at the U-matrix map (Fig.~\ref{fig:umatrix}).

Based on a visual inspection of the BMU map of variable sources (Fig. \ref{fig:bmu_var}), we can define some clear blobs as: \textit{Blob 1} is in the lower-right corner with coordinates $X \ga 35$, $Y \ga 10$, $Y \la 20$. The number of detections in this group is about $ 600$. \textit{Blob 2} is in the upper-right corner with coordinates $X \ga 35$, $Y \ga 20$. The number of detections in this group is about $1200$. This group shows a substructure in its lower part (blob 2b), with a separation at coordinates $X \ga 40$, $Y < 35$, containing about $ 250$ detections. \textit{Blob 3} is in the upper-left corner with coordinates $X \ga 5$, $X \la 20$, $Y \ga 33$. The number of detections in this group is about $ 250$. \textit{Blob 4} is in the upper center with coordinates $X \ga 25$, $X \la 35$, $Y \ga 35$ and the second one below at $X \ga 25$, $X \la 30$, $Y \ga 25$, $Y \la 35$. This latter contains about $ 50$ detections. \textit{Blob 5} is made out of the central group at $X \ga 15$, $X \la 32$, $Y \ga 12$, $Y \la 32,$ and contains about $ 550$ detections.\\

The $n_{var} = 2654$ detections have, on average, an order of magnitude higher signal to noise ratio (S/N)\footnote{Defined as the parameter PN\_8\_DET\_ML (maximum likelihood) in the 3XMM-DR4 Catalog.} than $n = 128\,925$ detections. This might be expected. Detections that are too faint do not cross the threshold of variability definition even if, intrinsically, they might be variable.

Training the SOM on just $n_{var} = 2654$ detections produces a uniform map without clearly separated blobs. It can be interpreted that one of the SOM results is finding and grouping interesting detections, and these detections mostly have a high S/N in order to be possible to distinguish their interesting features. In order for SOM or any other machine-learning algorithm to potentially "see" intrinsic features in faint sources (even if, for example, an astronomer with all the "ordinary" statistical tools would not be able to), instrumental and background effects would have to be input into the algorithm. Such an analysis is outside the scope of this paper.

\subsection{Classification of different groups} 
\label{res:quick_blobs}

\subsubsection{Quick look at all the blobs} 
\label{sec:groups}

In order to roughly examine each blob, we randomly chose a quarter of the $n_{var} = 2654$ detections and visually inspected their light curve to search for characteristic patterns. We divided the light curves into classes on the basis of their main shape: flares, bumps, multiple flares, multiple bumps, dips-eclipses, linear, and random.

\begin{figure}[htb]
    \centering
    \includegraphics[width=0.95\columnwidth]{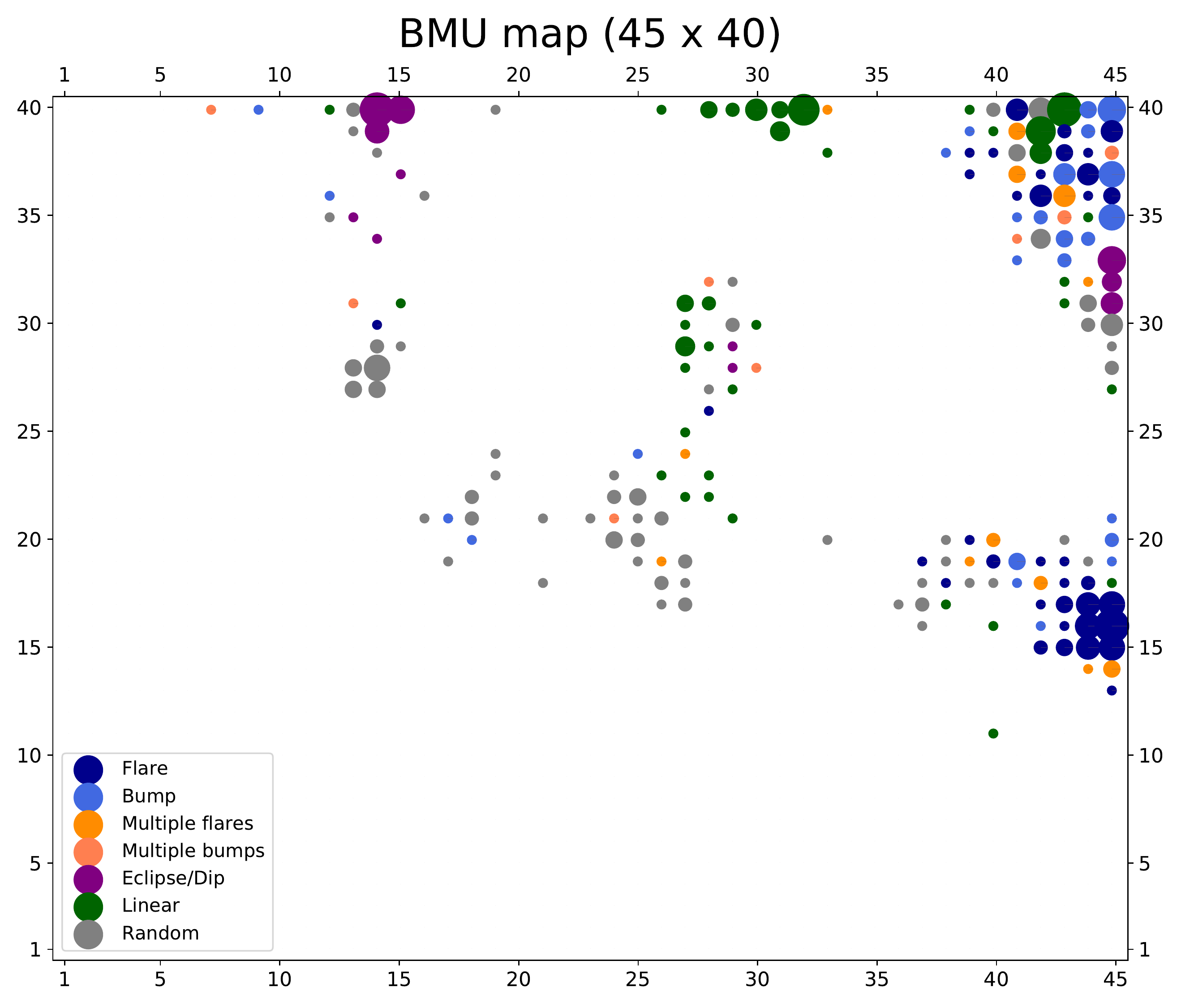}
    \caption{
    Classes of visually inspected variable detections. A quarter of $n_{var} = 2654$ was inspected and classes are presented as colored disks for each pixel. The most populated class at each pixel is shown. Flares are in dark blue, bumps in light blue, multiple flares in dark orange, multiple bumps in light orange, eclipses and dips in purple, linear in green, and random in gray. The size (area) of the disk corresponds to the number of detections belonging to the most populated class in a given BMU pixel.
    }
    \label{fig:bmu_var_vis_monopie}
\end{figure}

We classify any intense increase in flux followed by a fading to the quiescent level as flares and bumps, but, more specifically, flares follow a fast rise, exponential decay (FRED) time profile, while bumps have a more symmetrical shape (the same goes for multiple flares and multiple bumps). Although flares and bumps may originate from similar mechanisms (e.g., \cite{pye2015} show that coronal flares from stars may have comparable rise and decay times in a large fraction of cases), we decided to keep these two phenomenological classes of light curves  separate. Dips and eclipses are in a single class featuring any curve with one or more sudden and significant decrease in flux followed by a recovery to the upper level; the few cases of apparent dips and eclipses partially covered by the observation were treated case by case. The Random class includes light curves that do not show a distinct type of variability.
We built a map that shows the most numerous class in each pixel (Fig.~\ref{fig:bmu_var_vis_monopie}).

It is apparent that certain classes of light curves are predominantly concentrated in certain areas. For example, single flares are highly concentrated in the core of blob 1 (lower-right). Blob 2 (upper-right), the largest group, is composed of a variety of classes but dominated by multiple features; its substructure (the bottom part) is instead dominated by dips and eclipses. Blob 3 (upper-left) is mainly composed of dips and eclipses. Blob 4 (upper-center) is mainly composed of linear curves. The random curves are concentrated in blob 5 (central).

The SOM algorithm successfully extracted and grouped variable sources with the same variability behavior. Among the different blobs, the most intriguing from an astrophysical point of view are the ones dominated by flares, dips, and eclipses. For those groups, we extended our visual analysis.

\subsubsection{Single flares} 
\label{sec:flare}

From the analysis in Sect.~\ref{sec:groups}, we find that almost all of the flares are distributed over blob 1 and blob 2 with FRED-like flares being mostly present in the former. In blob 1, flares seem to be concentrated in the core, while in blob 2 their distribution is more complex and many other types of sources contribute to this blob.

Because of the large concentration of FRED-like flares, we examined blob 1  in detail. Due to the relatively large number of elements in blob 1, we visually examined only half of the approximately $650$ detections, with focus on phenomena that are likely to be related to an astrophysical flare.

\begin{figure}[ht]
    \centering
    \includegraphics[width=0.95\columnwidth]{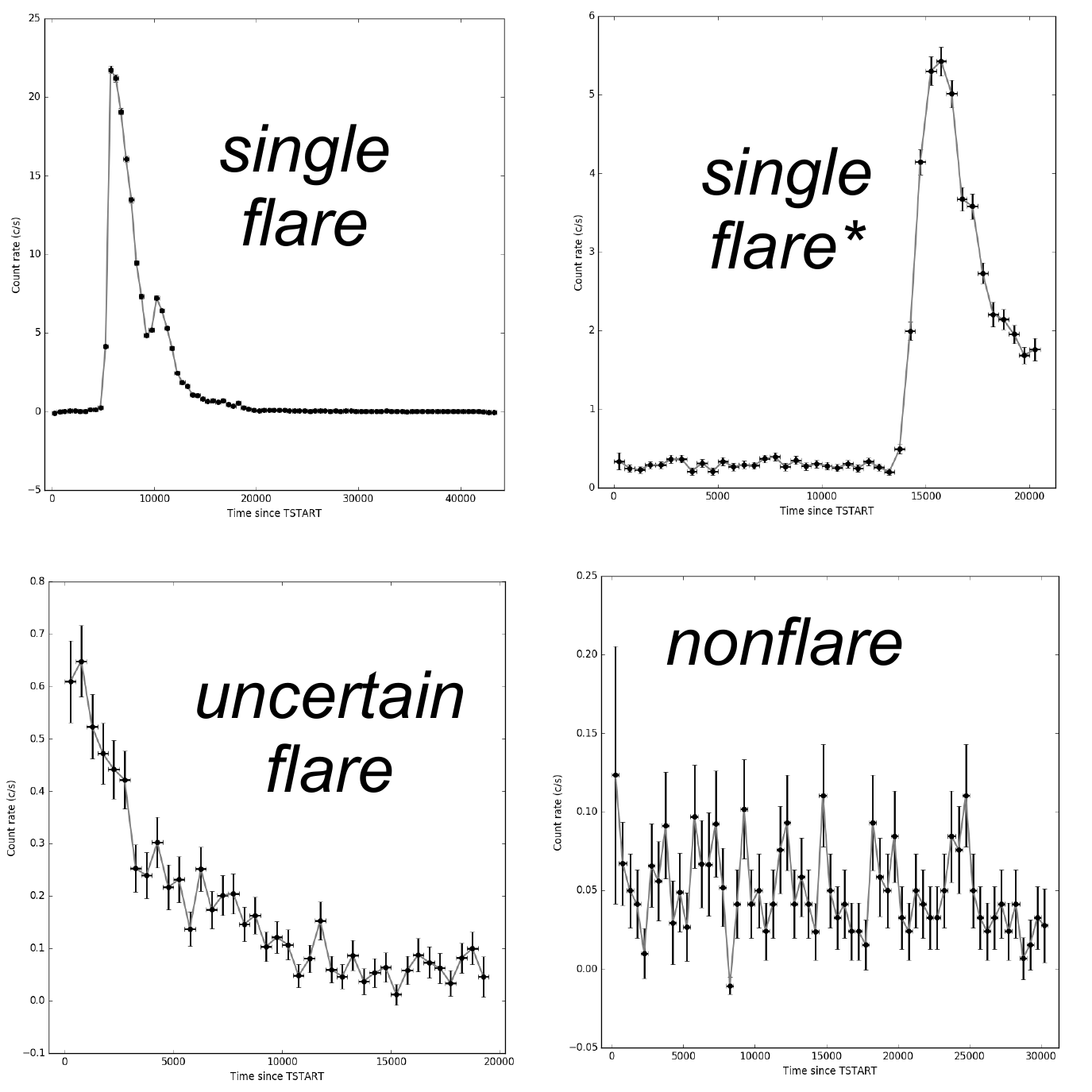}
    \caption{
    Examples of flares. The light curves are binned with 500 s time bins in one short-term exposure window; the vertical axis shows background-subtracted count rate. 
    (upper-left) Example of a bright flare: this detection is marked in the 3XMM-DR4 catalog as obs.id. 0604820301, src. 1. It is located at the BMU pixel X=45, Y=16.
    (upper-right) Example of a flare not fully covered by observations. This detection is marked in the 3XMM-DR4 catalog as: obs.id. 0134531601, src. 2. It is located at the BMU pixel X=45, Y=15.
    (bottom-left) Example of an uncertain flare. It shows exponential decay only, which could be the decaying part of the flare. This detection is marked in the 3XMM-DR4 catalog as: obs.id. 0302970201, src. 2. It is located at the BMU pixel X=45, Y=14.
    (bottom-right) Example of a nonflare. It shows a flickering behavior. This detection is marked in the 3XMM-DR4 catalog as: obs.id. 0302340101, src. 1. It is located at the BMU pixel X=39, Y=19.
    }
    \label{fig:flares}
\end{figure}

We defined three main classes of light curves:
(i) \textit{Single flare} -- the largest fraction are "textbook" flares with a FRED time profile fully within the observation period. Some are only partly within an observation period (e.g., with partial decay) and/or have a different time profile (e.g., "bumps" with similar rise and decay time). 
(ii) \textit{Uncertain} -- including all light curves showing some feature that could be related to an astrophysical flare (e.g., an exponential decay; a fast rise close to the end of the observation, etc.) but a different explanation could not be excluded.
(iii) \textit{Nonflares} -- including all light curves that did not have any relevant feature reminiscent of a flare.
Examples of single flares, uncertain flares, and nonflares are shown in Fig.~\ref{fig:flares}.

\begin{figure}[ht]
    \centering
    \includegraphics[width=0.95\columnwidth]{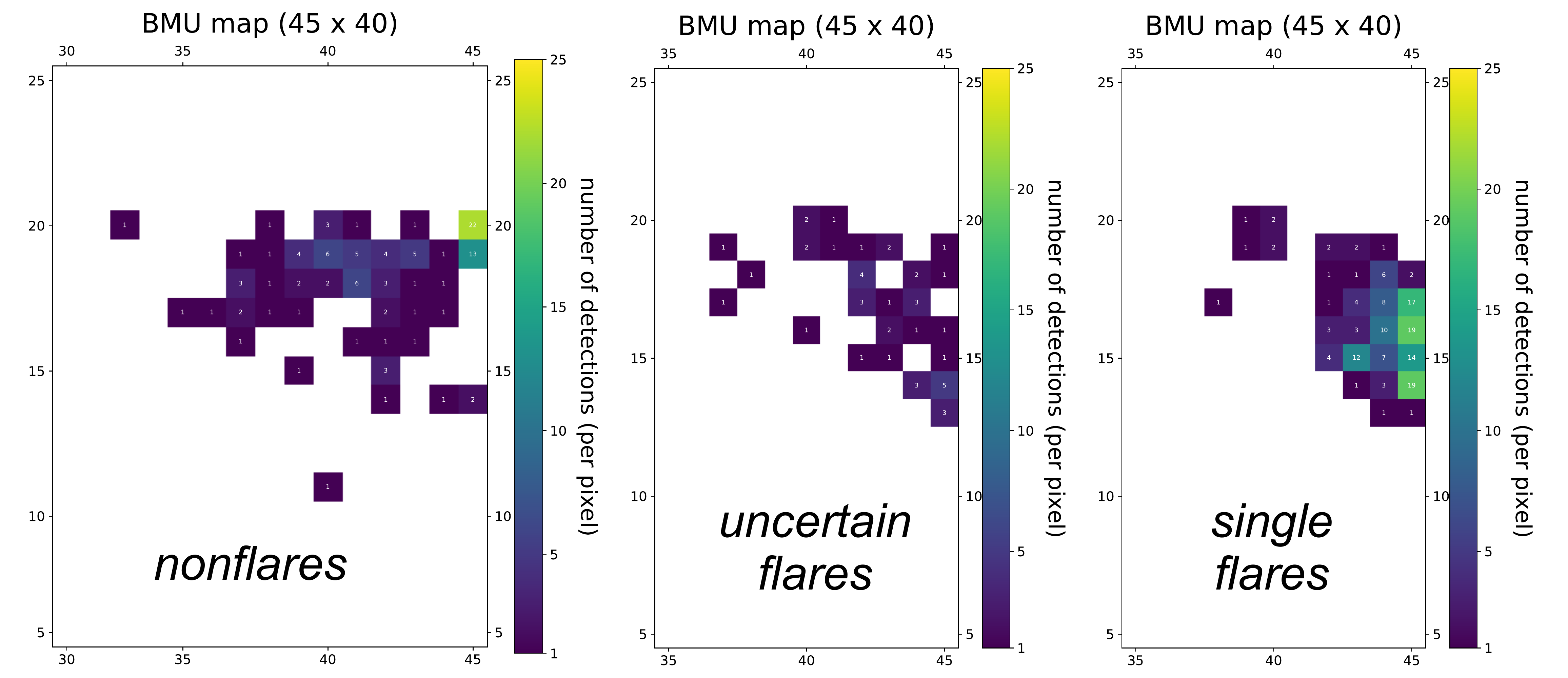}
    \caption{
    Distribution of visually inspected detections in blob 1. Single flares, uncertain flares, and nonflares are in the right, center, and left panels, respectively. All three panels have the same upper limit in the color bar for the purpose of direct comparison.
    }
    \label{fig:b1_f1}
\end{figure}

Figure~\ref{fig:b1_f1} shows the distributions of all classes. Single flares, uncertain flares, and nonflares are shown in the right, middle, and left panels, respectively. 

As can be seen, the concentration of clear flares is highest in the core of the blob, and gets diluted towards the blob edges. To crudely quantify this structure, the blob was divided into three parts: core, corona, and tail.

The concentration of single flares and other detections can be described as follows: The core is defined as pixels with coordinates $X \approx [15, 18],~Y = [44, 45]$. The core contains about $ 200$ detections of which 100 were visually inspected. Of these, 85 are single flares of which only five are bumps, 59 are FRED-like flares, and the rest are FRED-like flares not fully covered by observations. There are also 8 uncertain flares. The corona is defined as pixels with coordinates $X \approx [13, 20],~Y \geq 42$ excluding the core pixels. The corona contains about $ 300$ detections of which 150 were visually inspected. Of these, 59 (39\%) are single flares of which ten are bumps, 23 are FRED-like flares, and the rest are FRED-like flares that were not fully observed. There are also 27 (18\%) uncertain flares. The tail is defined as pixels with coordinates $X \approx [13, 20],~Y \leq 41$. The tail contains about $ 130$ detections, of which 65 were visually inspected. Of these, six (9\%) are single flares (two bumps and four faint FRED-like) and ten (15\%) are uncertain flares.

It is interesting to compare results of SOM with the results of the fit statistics from the flare model\footnote{The flare model in EXTraS catalog is defined as a constant plus fast rise and exponential decay (FRED).} from EXTraS.
We selected 566 detections (out of $n_{var} = 2\,654$) with a good flare model fit statistic\footnote{(a) the null hypothesis of the flare model is $<5\sigma$ and (b) an f-test confirms the statistical improvement by using the flare model instead of a constant at $>5\sigma$.}: these were visually inspected and classified into single flares, uncertain flares, and nonflares. Single flares make up 251 detections, uncertain flares 42, and non-flares 273: roughly half of the 566 detections selected based on a good flare model fit are not actually flares (e.g., the light curve has some random pattern that the automatic analysis managed to fit with the model). Thus, using an approach based on a model fitting, half of the flares are not genuine flares; with SOM, on the other hand, 93\% of the elements in the core of blob 1 are flares (or uncertain flares).

\begin{figure}[ht]
    \centering
    \includegraphics[width=0.95\columnwidth]{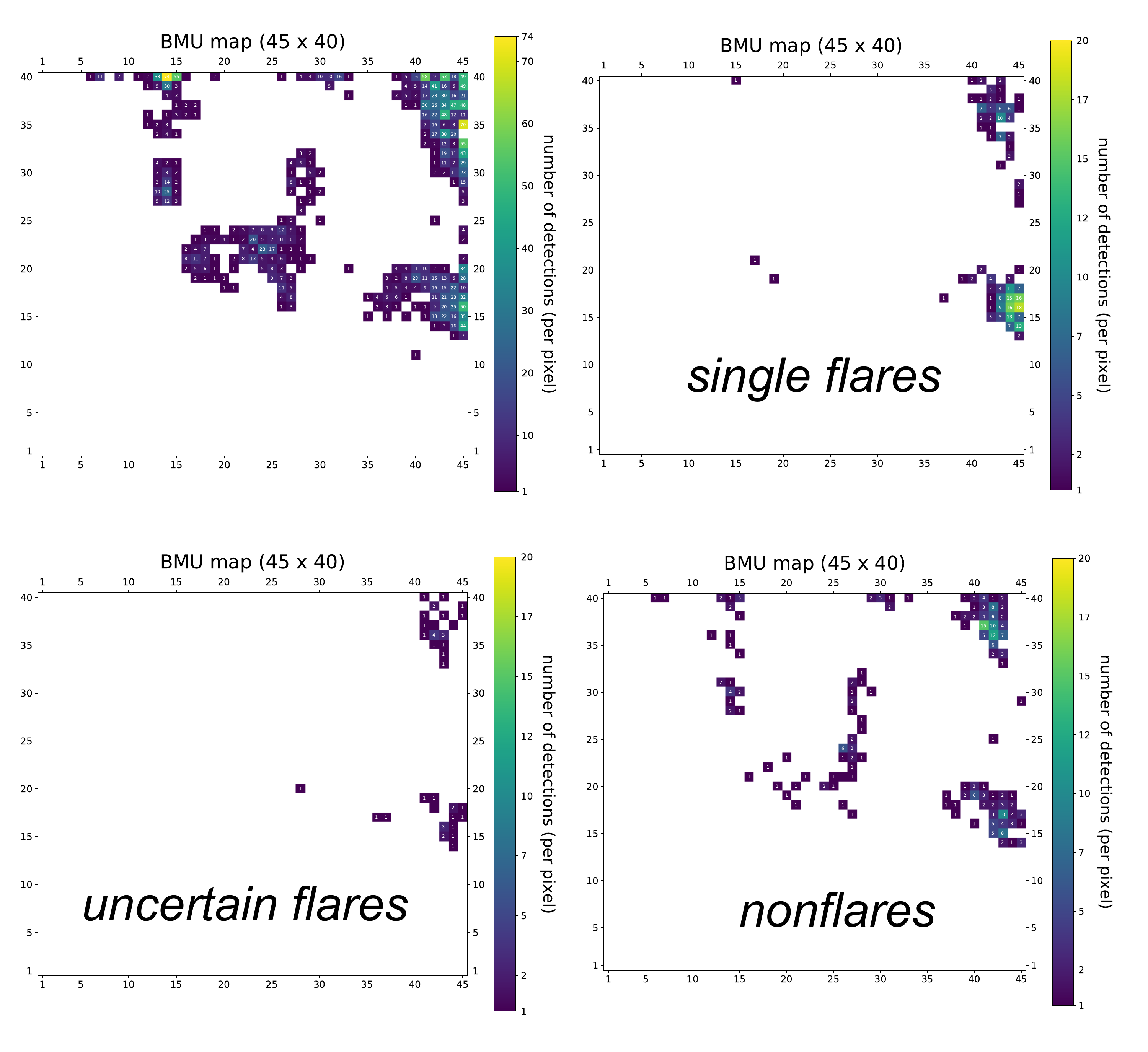}
    \caption{
    Visually inspected detections which have a good flare fit statistics. Anticlockwise from the upper-right, the panels show all $n_{var} = 2654$ variable detections, single flares (251 detections), uncertain flares (42), and nonflares (273).
    }
    \label{fig:fm}
\end{figure}

In Fig.~\ref{fig:fm} we show the three classes in the BMU map. About 90\% of the well-fitted, visually inspected flares fall into blob 1 and blob 2 (in a ratio of $\sim$ 2:1), in agreement with the findings of SOM (Figs.~\ref{fig:bmu_var_vis_monopie},~\ref{fig:b1_f1}). While flares are concentrated in the core in blob 1, they form a corona around the core  in blob 2.

On the other hand, only about $60\%$ of real single flares from the total visual inspection are well-fitted by the EXTraS flare model, either because the real flare is not a perfect FRED, the flare is superimposed on some other minor variations, or the fit fails.

We conclude that SOM was able to extract 97\% of the light curves with a "real" single flare and group them into two different groups (blob 1 and blob 2 with a ratio of $\sim$ 2:1). Within blob 1, flares compose up to 93\% of the core and up to 57\% of the corona; within blob 2, flares are concentrated in the corona. For comparison, through a classical model fitting analysis we are able to extract 60\% of the real single flares, and only 52\% of the well-fitted light curves contain a real single flare.\\

Most of the visually inspected flares are likely emitted by coronally active stars; this is either confirmed by the association of the flaring sources with stars in the Simbad database, or suggested by the soft spectrum of the X-ray sources and by their positional coincidence with cataloged optical/near-infrared objects. Peculiar phenomena of nonstellar origin can also be found in the sample: for instance, in the core of blob 1, we find the puzzling case of XMMU J134736.6+173403. This source is associated with a low-mass AGN and displays a sudden factor 6.5 decrease in flux occurring in about 1 hour\footnote{The overall shape of the light curve, featuring a "high state" lasting about 5 h, the sudden flux drop, and a "low state" lasting more than 10 h can be seen as a bump starting before the beginning of the observation.}. As discussed by \citet{carpano08} and \citet{carpano18}, this unusual drop in flux defies any easy explanation.

\subsubsection{Dips and eclipses} 
\label{sec:dip}

From a quick visual inspection of blobs (Sect.~\ref{sec:groups}), we find two distinct structures in the BMU map in which dips and eclipses are dominant: blob 3 and blob 2b (Fig.~\ref{fig:bmu_var}). These contain 38\% and 22\%, respectively, of the dips and eclipses found through the quick visual inspection; most of the remaining dips and eclipses are in the rest of blob 2 (23\%). The upper-left blob core is composed of pixels (14,40) and (15,40) and contains 129 light curves; it is surrounded by a corona of 87 light curves and a tail of 20 curves. The blob 2b core is composed of pixels (45,31), (45,32), and (45,33) just below, but separated from, the main structure of blob 2; it contains 127 light curves.

\begin{figure}[ht]
    \centering
    \includegraphics[width=0.95\columnwidth]{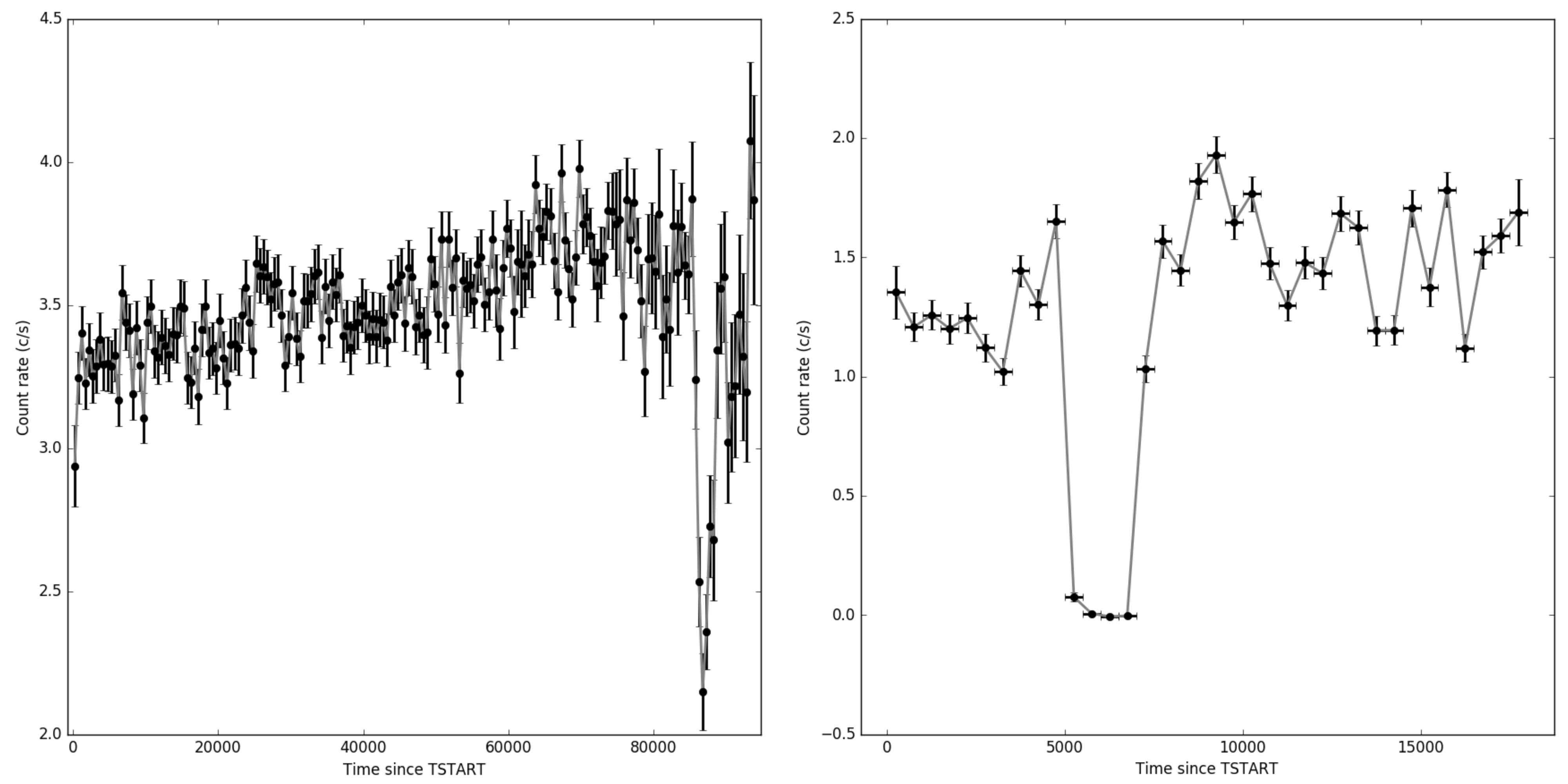}
    \caption{
    Example of a dip (left) and an eclipse (right). The light curves are binned with 500 s time bins in one short-term exposure window. The vertical axis shows the background-subtracted count rate. 
    The detection with the dip is marked in the 3XMM-DR4 catalog as obs.id. 0200470101, src. 1. It is located at the BMU pixel X=14, Y=40.
    The detection with the eclipse is marked in the 3XMM-DR4 catalog as obs.id. 0110660101, src. 1. It is located at the BMU pixel X=45, Y=32.
    }
    \label{fig:dip-ecl}
\end{figure}

Here, we examined blob 3 and blob 2b  in detail. We visually inspected all the light curves in these regions  in
detail, focusing on phenomena that are likely to be related to a dip or an eclipse. We divided the sources into three classes: random, dip, and eclipse. Random light curves do not show any apparent fall--rise behavior (even if they can show any other behavior described in Sect.~\ref{sec:groups}). We classified the remaining light curves as "dip" or "eclipse" based on the literature for associated sources. If there was no association with a dipping or eclipsing source, classification was based on the shape of the fall and rise: dips are short (less than 5 bins) with a clear decrease and increase (typically a "V" shape), while eclipses are longer and/or characterized by a constant, low flux level (typically a "U" shape). An example of a dip and an eclipse in shown in Fig.~\ref{fig:dip-ecl}.

We find that, in the core of blob 3, 90\% of the light curves are dips or eclipses, while in the corona this percentage is 45\%, and in the tail these represent  20\%. Most of them (97\% in the core, 80\% in the corona, and 83\% in the tail) are dips. While some of the dips are instrumental errors occurring at the beginning or the end of the observation, we find many well-known dipping sources, for example: 2XMM J125048.6+410743 \citep{2013ApJ...779..149L} and 3XMM J004232.1+411314 \citep{2017ApJ...851L..27M}.

In the core of blob 2b. 60\% of the light curves are dips or eclipses, of which 60\% are eclipses and the remaining dips are usually longer than those in blob 3 and/or their statistics are poorer. Among them, we find many well-known eclipsing sources, for example: V* V1727 Cyg \citep{2007AA...476..301B} and V* XY Ari \citep{2007AA...472..225N}. We investigated the existence of a corona in the lower-left part of blob2, but only 8 of 105 light curves show a clear eclipse or dip.

The SOM algorithm was therefore able to extract 83\% of the light curves that show one or more dip or eclipse and to group them into blob 3 and blob 2 (quick visual analysis in Sect.~\ref{sec:groups}). From the detailed visual analysis, we find that, within blob 3, dips vastly dominate over eclipses and dips and eclipses compose 90\% of the blob core and 45\% of its corona. Blob 2b is dominated by dips and eclipses of which 60\% are single or multiple eclipses while most of the dips are wider than the ones in blob 3. In the core of this blob, 60\% of the light curves show one or more dips or eclipses.

In order to confirm and compare the results based on the visual inspection, we cannot rely on the eclipse model from EXTraS as we did for flares (Sect.~\ref{sec:flare}). The eclipse model is indeed quite simple, with a perfect U shape, and thus it cannot describe more complex light curves (e.g., with a rise and decay time), dips, or periodic features; moreover, the eclipse model usually fits most of the random increases or decreases of a low-statistics light curve well. A rough comparison comes from the sample used for the quick visual analysis in Sect.~\ref{sec:groups}: the number of well-fitted eclipses\footnote{We use the same definition as in Sect.~\ref{sec:flare}.} is more than twice the number expected from the visual inspection, while only half of the dips and eclipses from visual inspection are well fitted by the eclipse model. Instead, we randomly selected a number of X-ray eclipsing-like sources observed by {\it XMM-Newton} from  the literature. Our random selection comprises different types of objects, with one or more observations, and with one or more features in the same exposure. We selected 12 sources for a total of 22 detections (see Table~\ref{tab:literature} in Appendix~\ref{app:res}). Of the 22 detections, 16(+1) fall in the core (corona) of blob 2b. Four detections fall in the remaining part of blob 2, but always at X = 45. One detection falls in blob 3. It is also interesting to note that different exposures of the same source usually fall in the same or the adjacent pixel.

\subsection{Interesting sources}
\label{sec:new}

Dips and eclipses are quite rare and are interesting from an astrophysical point of view because they usually indicate binary systems and/or imply the presence of an accretion disk or  blobs of dust. In this case, the SOM is particularly useful for the discovery of single, interesting systems. Therefore, we searched our sample for unpublished features and obtained eight sources.
In the following, we report a brief description of them, including their {\it XMM-Newton} names, source numbers, and coordinates (all of which come from the 3XMM-DR4 catalog).

{\it 3XMM J063736.4+053932} (obs.id. 0655560101, src. 1, BMU pixel 14,39) is located at RA(J2000) 06:37:36.48, Dec(J2000) +05:39:32.59. The EXTraS pn light curve shows a total eclipse in the last 2 ks (over a 26 ks exposure) not covered by the MOS cameras. The positional coincidence with the 8.5 V magnitude star HD 47179 suggests this source is a stellar binary system.

{\it 3XMM J081928.9+704219} (obs.id. 0200470101, src. 1, BMU pixel 14,40) is located at RA(J2000) 08:19:29.00 Dec(J2000) +70:42:19.17. The EXTraS pn light curve clearly shows a dip that halves the X-ray count rate (5 ks over a 83 ks exposure). It falls during a very high-background period but the dip shape does not seem to be correlated with the background. This source is associated with the well-studied ultraluminous X-ray source Holmberg II X-1. \citet{2006MNRAS.365..191G} analyzed this detection, but the time of the dip was discarded because of the high background. Although EXTraS tools are well suited to deal with high background \citep{2021A&A...650A.167D}, a dedicated analysis is required to confirm this feature.

{\it 3XMM J133000.9+471343} (obs.id. 0303420201, src. 2, BMU pixel 14,40) is located at RA(J2000) 13:30:00.96 Dec(J2000) +47:13:43.65. The EXTraS pn light curve shows a peculiar flickering pattern, possibly quasi-periodic, with a timescale of $\sim$ 20 min. Light curves from MOS cameras confirm this peculiar variability. Interestingly, the source is M51 ULX-7, a pulsating ($\sim$ 2.8 s) ultraluminous X-ray source with an orbital period of $\sim$ 2 days and a possible super-orbital modulation of $\sim$ 38.9 days \citep{2020ApJ...895...60R,2021AAS...23822502V}.

{\it 3XMM J031822.1-663603} (obs.id. 0405090101, src. 2, BMU pixel 14,40) is located at RA(J2000) 03:18:22.17 Dec(J2000) -66:36:03.4. The EXTraS pn light curve shows a random variability with an eclipse-like sudden drop ($\sim$ 40\% of the average count rate) during the last 3 ks of the observation. This drop is confirmed by both MOS cameras. This source is associated with the pulsating ($\sim$ 1.5 s) ultraluminous X-ray source NGC1313 X-2 \citep{2019MNRAS.488L..35S,2021AA...652A.118R}.

{\it 3XMM J080945.3-472110} (obs.id. 0112670501, src. 4, BMU pixel 45,32) is located at RA(J2000) 08:09:45.35 Dec(J2000) -47:21:10.16. The EXTraS pn light curve starts in a constant, low state that lasts for 3ks (over a 28 ks exposure) and then suddenly rises by a factor of $\sim$ 10 in count rate. It can be interpreted as either an eclipse or a FRED flare with a very long characteristic decay time ($\sim$ 30 ks). Data from MOS cameras are not available. We note that the only other {\it XMM-Newton} observation (55 ks exposure) of this source shows a count rate compatible with the low state. This source is positionally consistent with the young stellar object candidate 2MASS J08094536-4721101. 

{\it 3XMM J063045.4-603113} (obs.id. 0679381201, src. 1, BMU pixel 45,32) is located at RA(J2000) 06:30:45.42 Dec(J2000) -60:31:13.15. The EXTraS light curve shows an eclipse or a series of dips in the last 3ks of the observation (over a 13ks exposure), with a drop of $\sim$ 75\% of the count rate. This behavior is confirmed by both MOS cameras. The source is associated with XMMSL1 J063045.9-603110, a peculiar transient source \citep{2011ATel.3811....1R} proposed to be  a tidal disruption event \citep{2016AA...592A..41M}, but later spectroscopically classified as a nova \citep{2017AJ....153..144O}.

{\it 3XMM J182422.8-301833} (obs.id. 0551340201, src. 52, BMU pixel 45,32) is located at RA(J2000) 18:24:22.82 Dec(J2000) -30:18:33.2. The EXTraS pn light curve clearly shows a periodic, possibly sinusoidal (or, a series of dips)  pattern. Indeed, the search for periodic sources performed within EXTraS reveals a significant coherent signal at 2919 s (a complete analysis will be presented in the EXTraS pulsators catalog (Israel et al. in preparation). The X-ray source has a few possible optical counterparts and is also positionally consistent with a WISE source. It could be a low-mass X-ray binary, but a dedicated analysis is needed to confirm this hypothesis.

{\it 3XMM J053427.3-052420} (obs.id. 0403200101, src. 5, BMU pixel 45,33) is located at RA(J2000) 05:34:27.37 Dec(J2000) -05:24:20.92. The EXTraS pn light curve starts in a constant, low state that lasts for 20 ks (over a 90 ks exposure) and then suddenly rises by a factor of $\sim$ 2 in count rate. This should be interpreted as an eclipse -- a FRED flare would have a very long characteristic decay time ($\sim$ 90 ks). Data from MOS cameras confirm the variability pattern. Other {\it XMM-Newton} observations see the source ---which is usually variable--- in different states. This source is positionally consistent with the 12.4 V magnitude variable star of Orion type V* V1961 Ori.

\subsection{Caveats and robustness checks}
\label{res:caveats}

In general, a SOM is not guaranteed to correctly represent all the relevant structure of a data set. A simple check of whether the training process led to an acceptable result is to consider the distribution of each variable from the initial high-dimensional space: is it the same on the SOM neurons as in the original data? As the aim of training a SOM is for the neurons to behave like representative data points or prototypes, this is clearly a minimum requirement. If the neurons have a very different distribution with respect to the original data then training did not work as expected, perhaps having to few iterations. In Fig.~\ref{fig:som-units_x31} we show all 31 normalized parameters, distinguishing all the $128\,925$ detections, the 2654 variable detections, and the 1800 SOM neurons. It is clear that our SOM neurons generally follow the distribution of the original data on each parameter.

\begin{figure}[ht]
    \centering
    \includegraphics[width=0.95\columnwidth]{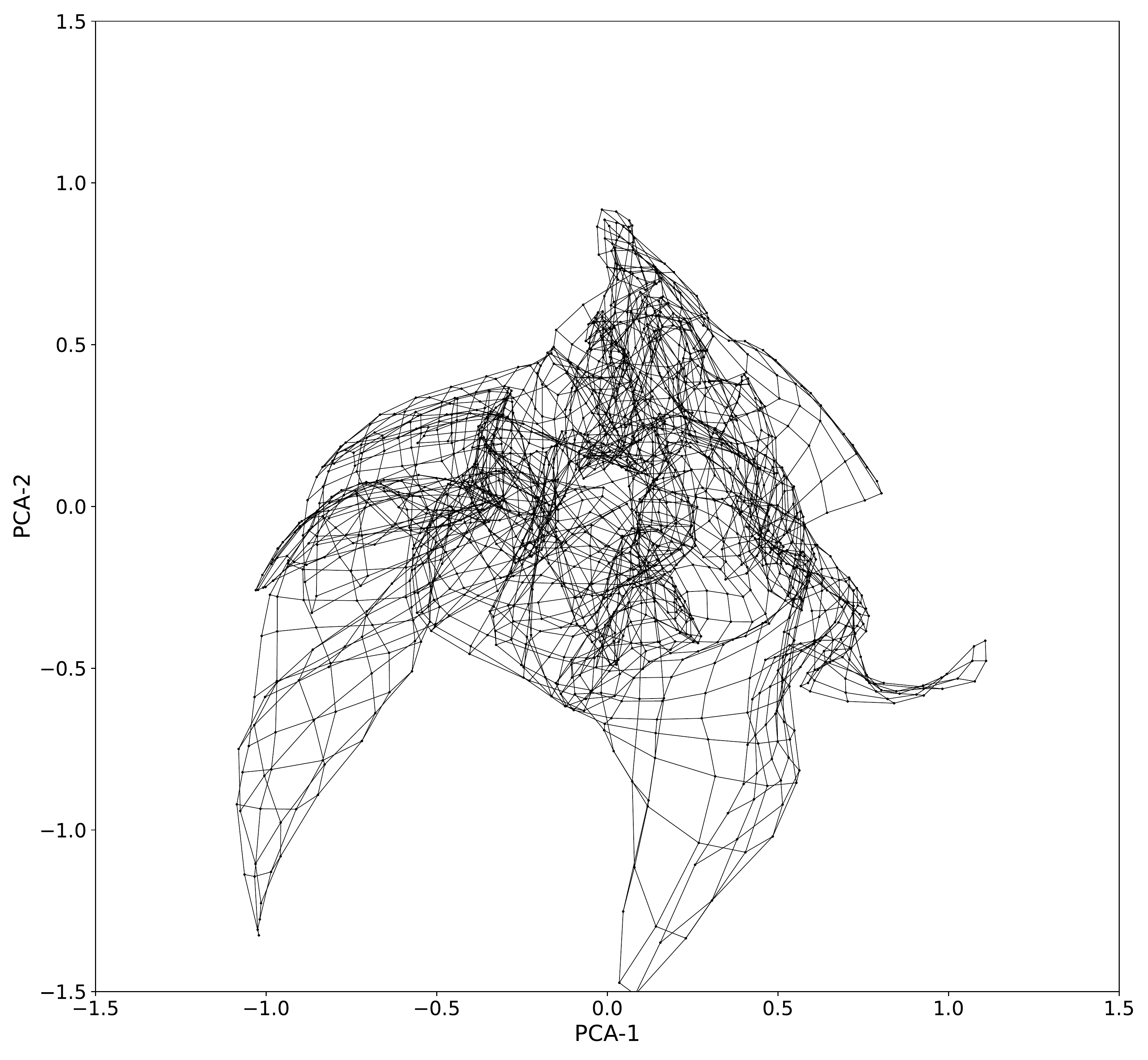}
    \caption{
    SOM map projected on a plane formed by the two largest PCA vectors. Dots represent SOM neurons while lines are connections between neighboring neurons.
    The evolution of the SOM map projection during training is available as an online movie `fig\_s5\_som\_pca12\_movie'. One can see how the map goes from its starting position as a rectangle, changes during rough training (1-80 epoch), and converges to its final position during fine training (81-160 epoch). 
    }
    \label{fig:s5:som_pca}
\end{figure}

Another test with a similar goal is to compare the results of our SOM to those of other, simpler dimensionality-reduction approaches. The simplest is PCA, which is a linear procedure building a set of orthogonal variance-maximizing linear combinations of the (standardized) original coordinates. Retaining only the first two PCA coordinates ---which explain the most variance in the data set--- allows visualization on a plane. However, the linear nature of PCA makes it hard for it to correctly represent nonlinear structure. In Fig.~\ref{fig:s5:som_pca}, our SOM map is projected from the original 31-dimensional parameter space on to a plane formed by the first two PCA coordinates. Our map can clearly be seen to generally cover this PCA plane, even though it is twisted in a nontrivial way. This suggests that the original parameters are related in complex nonlinear ways, justifying the need for a SOM, or for nonlinear dimensionality reduction in general, as opposed to PCA. A possible cause for concern is that the SOM may have a complex shape (Fig.~\ref{fig:s5:som_pca}) because it is trying to compensate for the difference between the intrinsic dimension of the data set and the map intrinsic dimension of $D = 2$. Increasing the dimensionality of our SOM by arranging its neurons on a lattice in three-dimensional space would address this issue but make visualization more cumbersome. We therefore chose not to explore this option in the current paper, even though it may be worth investigating in a subsequent one.

\section{Conclusions}
\label{conclusions}

The {\it XMM-Newton} telescope greatly advanced our knowledge of the X-ray sky, with the EXTraS project detecting and characterizing the time variability of over 300\,000 sources. The resulting data set poses the typical challenges of big data, serving as a clear illustration that X-ray astronomy is transitioning into this regime. In this context, traditional approaches (e.g., human visual inspection) do not allow us to take full advantage of the opportunities offered by the data.

In this paper we applied a machine learning approach with the goal of automatically organizing data to maximize the effectiveness of direct human inspection. To this end, we selected a subset of parameters ---from the originally large number provided by EXTraS--- that characterize the variability of each source, and applied dimensionality reduction to the resulting data set. This was achieved using the SOM algorithm, which represents the data on a plane, attempting to respect the topology of the original high-dimensional space. By construction, the SOM builds a grid of representative points that summarize the original data, and lays them out grouped together based on the similarity of their characteristics. It thus clusters the data while reducing its dimension to a plane for visualization purposes. This is something that would not be achieved by a linear approach such as PCA, which would miss most of the intrinsically nonlinear structure of our data set that SOM is able to capture, as shown in Fig.~\ref{fig:s5:som_pca}.

Despite being a time-tested algorithm which has already been used in astronomy, this is the first time\footnote{As far as we can tell.} SOM is applied in this context (large X-ray data set). As a result, we streamlined a process of source recognition that would otherwise have been driven by serendipitous discovery, finding flares, dips, eclipses, and other source types, all arranged into contiguous clumps in the SOM plane. Used in this way, SOM allows an astronomer to concentrate on inspecting regions of data space that appear scientifically promising.

We highlighted the problem of straightforward temporal model fitting to light curves and its use to characterize them, especially when data are noisy, and showed that the SOM algorithm can overcome this problem to an extent by utilizing parameters derived from the light curves.

With the introduction of this new tool, we were able to explore the EXTraS data set, focusing on variable sources, quickly selecting a number of objects that have interesting properties that warrant further investigation, including different kinds of binary systems (from binary stars to ULXs) as well as more peculiar sources. While some of these objects were already investigated and described in the literature, for example the most luminous dipper known 3XMM J004232.2+411314 \citep{2017ApJ...851L..27M}, the peculiar transient 3XMM J063045.4-603113 \citep{2016AA...592A..41M}, and the poorly understood, low-mass AGN XMMU J134736.6+173403 \citep{carpano08}, we also extracted some new interesting sources (Sect.~\ref{sec:new}). It should be noted that this data set, based on observations collected until 2012, was widely analyzed by the astronomical community for years before this work.

Summarizing data becomes more and more valuable as data sets grow. Our approach is therefore promising, especially in the light of the upcoming new EXTraS data, not to mention future space missions that may yield much richer and sensitive data than {\it XMM-Newton}, such as the ESA ATHENA observatory. Furthermore, our results pave the way for upcoming work focused on supervised learning, where the goal is to look for specific objects (e.g., "FRED" flare-like events or eclipses) armed with a good understanding of the parameter space. This will allow us, for instance, to visualize the predicted classification of a supervised learner on the SOM plane, which is an effective interpretability technique \citep[see e.g.,][]{molnar2019}.


\begin{acknowledgement}
M. K. acknowledges financial support from the Italian Space Agency (ASI) under the ASI-INAF agreement 2017-14-H.0.~
M. P. acknowledges financial support from the European Union’s Horizon 2020 research and innovation program under the Marie Sklodowska-Curie grant agreement No. $896248$.~
All the authors acknowledge comments and suggestions from the referee that led to the improvement of this work.
\end{acknowledgement}


\bibliographystyle{aa} 
\bibliography{main}


\begin{appendix}
\label{apdx}

\section{Data selection, normalization, and correlation}
\label{app:data}

\subsection{Data normalization}
\label{app:datanorm}

The distribution of the values of each one of the 31 parameters we selected is presented in Fig.~\ref{fig:params1}. Most of the parameters are distributed with very narrow cores centered on zero and long tails either in both positive and negative directions or in the positive direction only. For some parameters, such as kurtosis and skewness, their distribution is highly asymmetric. In such cases histograms were binned in a symmetric logarithmic scale centered on zero and a linear scale around zero in order to have a clearer idea of their distribution. The parameter groups CDF\_TFRAC\_* and CDF\_RFRAC\_* span between zero and one, and their distribution does not have such a narrow core compared to the tail. They were plotted with linear time bins. All the parameters have exactly the same number of values. This is necessary for the SOM algorithm to work, that is, each detection in the parameter space has all its 31 coordinates defined. 

All three p-values (histograms 1, 2, and 4  in Fig.~\ref{fig:params1}) were recalculated with higher precision and converted into one-sided sigma values in such a way that higher sigma corresponds to a poorer fit of the model. In this way, sigma is a proxy for variability against the three models. Another reason to transform p-values is that the vast majority of them are concentrated towards zero and one, and are hardly distinguishable in a linear scale; however they correspond to very different levels of goodness of fit to their models. Even with this higher precision recalculation, many values are capped at $\sim 37~\sigma$ and so they fall in the final bin.

Parameters in the group CDF\_RFRAC\_* show spikes on top of a smooth distribution. The reason for this is that they are defined as percentage of time the source spends in a certain state, and as there is always a finite number of time bins, this introduces a form of discretization.\\

The SOM algorithm typically relies on Euclidean distance in parameter space to quantify the dissimilarity between data points. To avoid over- or under-weighting parameters based on their units of measure, their values have to be normalized on a similar scale in order to give each of them similar influence in guiding the SOM training process.

Simply normalizing to a fixed range by linearly rescaling has several drawbacks in the presence of long tails and/or outliers. This prevents us from simply assigning the minimum and the maximum of each variable for instance to [0, 1], as most values would end up concentrated around zero. Similar concerns also prevent us from normalizing by setting the sample standard deviation to unity.

We quantify the importance of parameter distribution core, tail and outliers by taking the ratio of the standard deviation to median absolute deviation, $r_{nrw}$ (third column in Table~\ref{tab:params}). The median absolute deviation is robust to long tails and outliers, while the standard deviation is not, and so large values of $r_{nrw}$ imply the presence of long tails and/or outliers. About half of parameters have $r_{nrw} \gg 1$ with skewness and kurtosis having $r_{nrw} \sim 10^7$ and $r_{nrw} \sim 10^{10}$. The other half have $r_{nrw} > 1$, $r_{nrw} \gse 1$ or $r_{nrw} \lse 1$.

To solve this issue we relied on a power transform of the affected variables. A power-law exponent $p_{nrw}$ was defined as $p_{nrw} = \tfrac{1}{ \log{(10 \times r_{nrw})}}$. The idea is that $p_{nrw}$ decreases slowly from 1 with increasing $r_{nrw}$ and when $r_{nrw} = 1$, $p_{nrw} = 1$ ($p_{nrw}$ was set to 1 also when $r_{nrw} \lse 1$). Therefore for all parameters $p_{nrw} \le 1$ (and positive).

For each set of parameter values, the distance between two successive values $\Delta x$ was transformed as $(\Delta x)^{p_{nrw}}$. This has the effect of increasing the distance between values which are too close and decreasing the distance between values which are too distant. Also, the effect of increasing or decreasing distance is larger (lower $p_{nrw}$) if the parameter has higher $r_{nrw}$. Crudely speaking, this process stretches the cores and squeezes the tails with an intensity depending on the initial distribution. This preserves the ordering (ranking) of values. Finally, all transformed parameters were rescaled linearly to the range [0, 1]. 

The normalized distribution of each parameter is shown in Fig.~\ref{fig:params2}. All histograms are binned linearly between zero and one. The normalized values of the parameters are filling up the same range of [0, 1], and are much more evenly distributed than the original values, while maintaining the general shape of the original distribution. Parameters with $r_{nrw} \lse 1$ ($p_{nrw} = 1$) have an identical distribution before and after normalization; parameters with $r_{nrw} > 1$ ($p_{nrw} \lse 1$) have a similar distribution in the two cases; the distribution of parameters with $r_{nrw} \gg 1$ ($p_{nrw} < 1$) is the most affected by normalization (in the sense of core stretching and tail squeezing). The extreme values (i.e., potential outliers) are still at the edges of their distribution, but are not too far from the majority of values.

\subsection{Data correlation}
\label{app:datacorr}

\begin{figure*}[htb]
    \centering
    \includegraphics[width=0.95\textwidth]{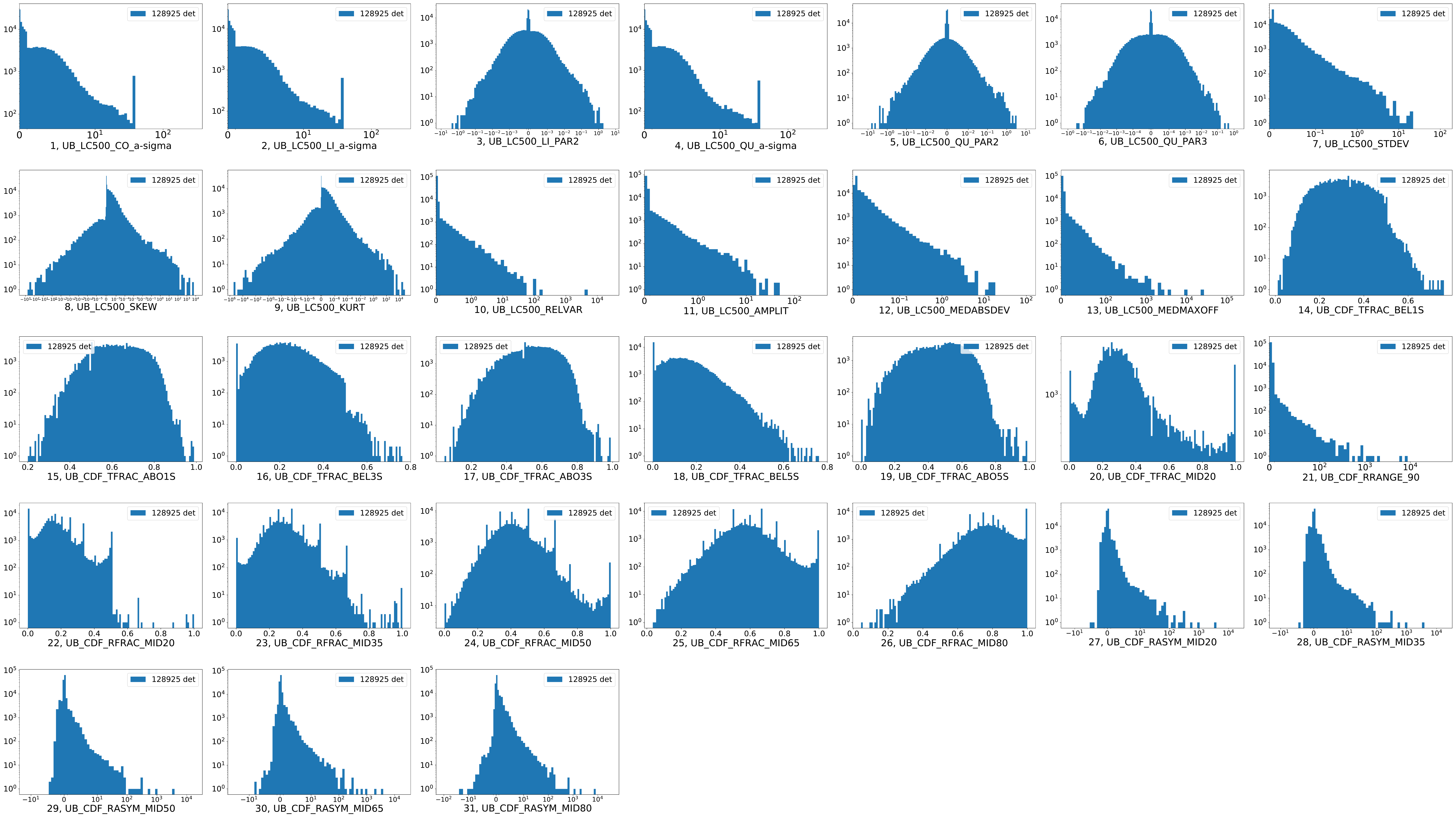}
    \caption{
    Histograms of the $m = 31$ parameter distributions. Each parameter is numbered corresponding to the order in Table~\ref{tab:params}. The number of detections is the same for each parameter and is shown in the legend in every histogram.
    Histogram binning adaptively switches between linear (around zero) and logarithmic (in the distribution tails) in most cases to best present the distribution of each parameter. Number labels were omitted from ticks near zero for clarity. The vertical axes are in logarithmic scale.
    }
    \label{fig:params1}
\end{figure*}

\begin{figure*}[htb]
    \centering
    \includegraphics[width=0.95\textwidth]{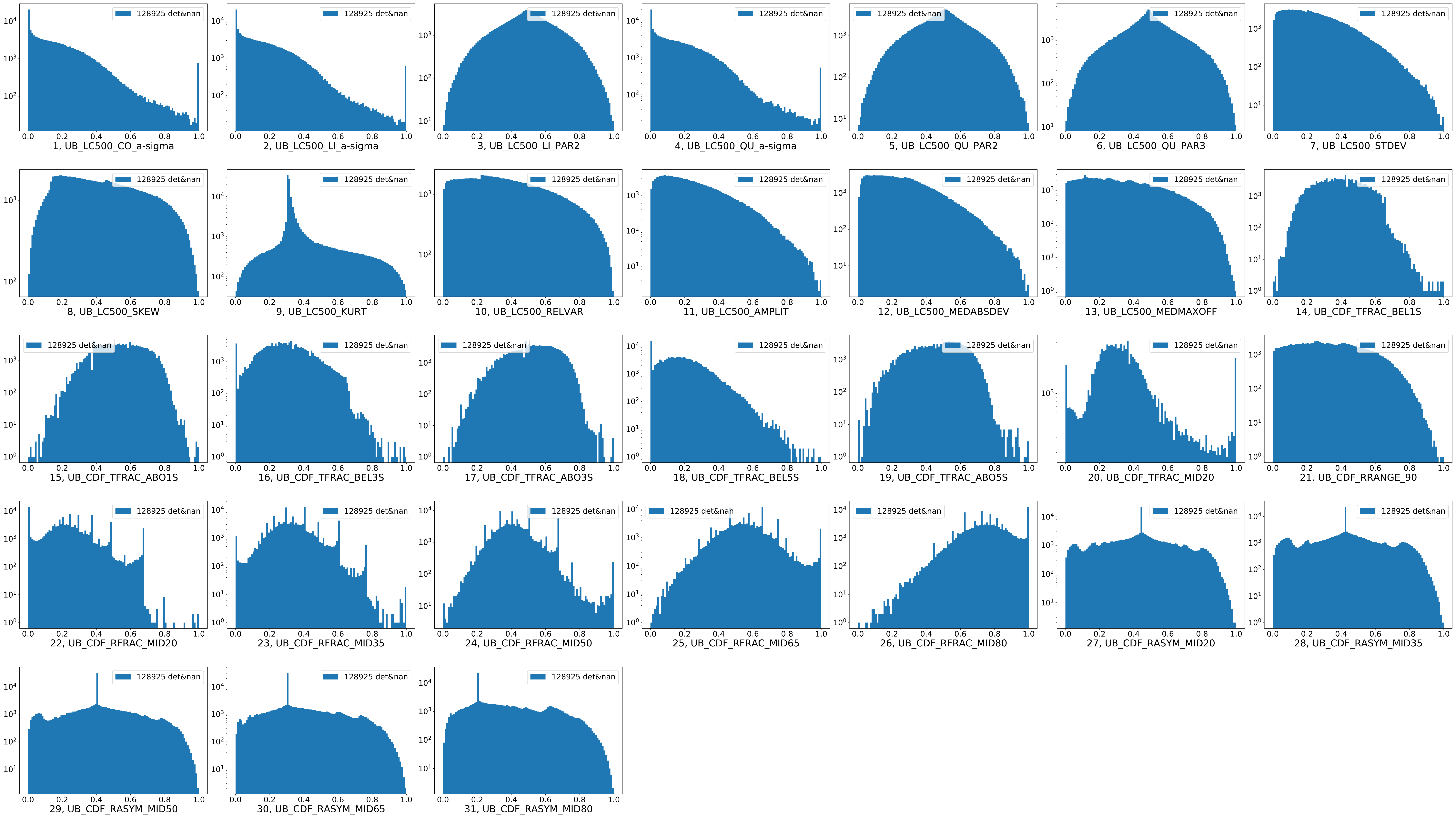}
    \caption{
    Histograms of $m = 31$ normalized parameter distributions. Each parameter is numbered corresponding to the order in Table~\ref{tab:params} and Fig.~\ref{fig:params1}. The number of detections is the same for each parameter and is shown in the legend in every histogram. All histograms are binned in linear scale ranging from zero to one. The vertical axes are in logarithmic scale, as in Fig.~\ref{fig:params1}.
    }
    \label{fig:params2}
\end{figure*}

As can be seen from histograms (Figs.~\ref{fig:params1}, ~\ref{fig:params2}) several parameters appear to share a similar distribution. We quantified their pairwise correlations by calculating the "Pearson r" correlation coefficient, which measures the linear correlation between parameters. 

\begin{figure}[htb]
    \centering
    \includegraphics[width=0.95\columnwidth]{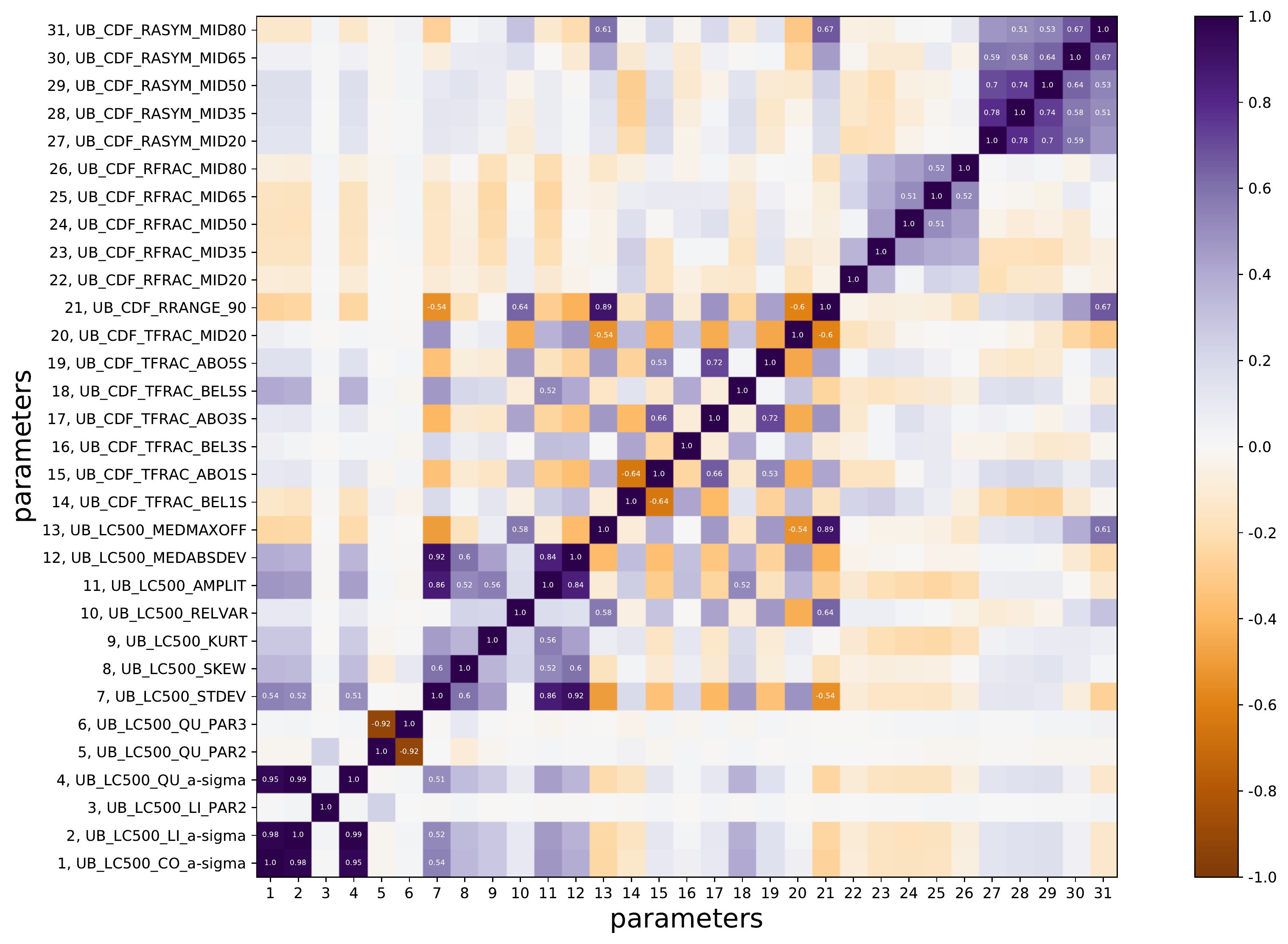}
    \caption{
    Pearson r correlation matrix of all $m = 31$ normalized parameter values. Positive (purple) values correspond to positive correlation while negative (brown) values to negative correlation. Correlation coefficients with absolute value less than 0.5 are not explicitly written.}
    \label{fig:corr}
\end{figure}

The correlation matrix for our 31 normalized parameters is shown in Fig.~\ref{fig:corr}. As the distribution of parameter values is featured, "Kendall rank $\tau$" and "Spearman rank $\rho$" correlation coefficients\footnote{Rank correlation coefficients compare two distributions based on the ordering of their values (from smallest to largest), not on the values themselves. As long as ordering is the same, the distribution of values is not important.} were checked. They are similar to Pearson r coefficients.\\

As can be seen from Fig.~\ref{fig:corr} there are many parameters with a large $|r| > 0.5$ association between each other. As expected some parameters form groups with high (anti)correlation such as the three UB\_CDF\_TFRAC\_ABO*S parameters, the two UB\_LC500\_QU\_PAR*, both standard deviations (UB\_LC500\_STDEV and UB\_LC500\_MEDABSDEV), the five UB\_CDF\_RASYM\_MID* and others.

\begin{figure}[htb]
    \centering
    \includegraphics[width=0.8\columnwidth]{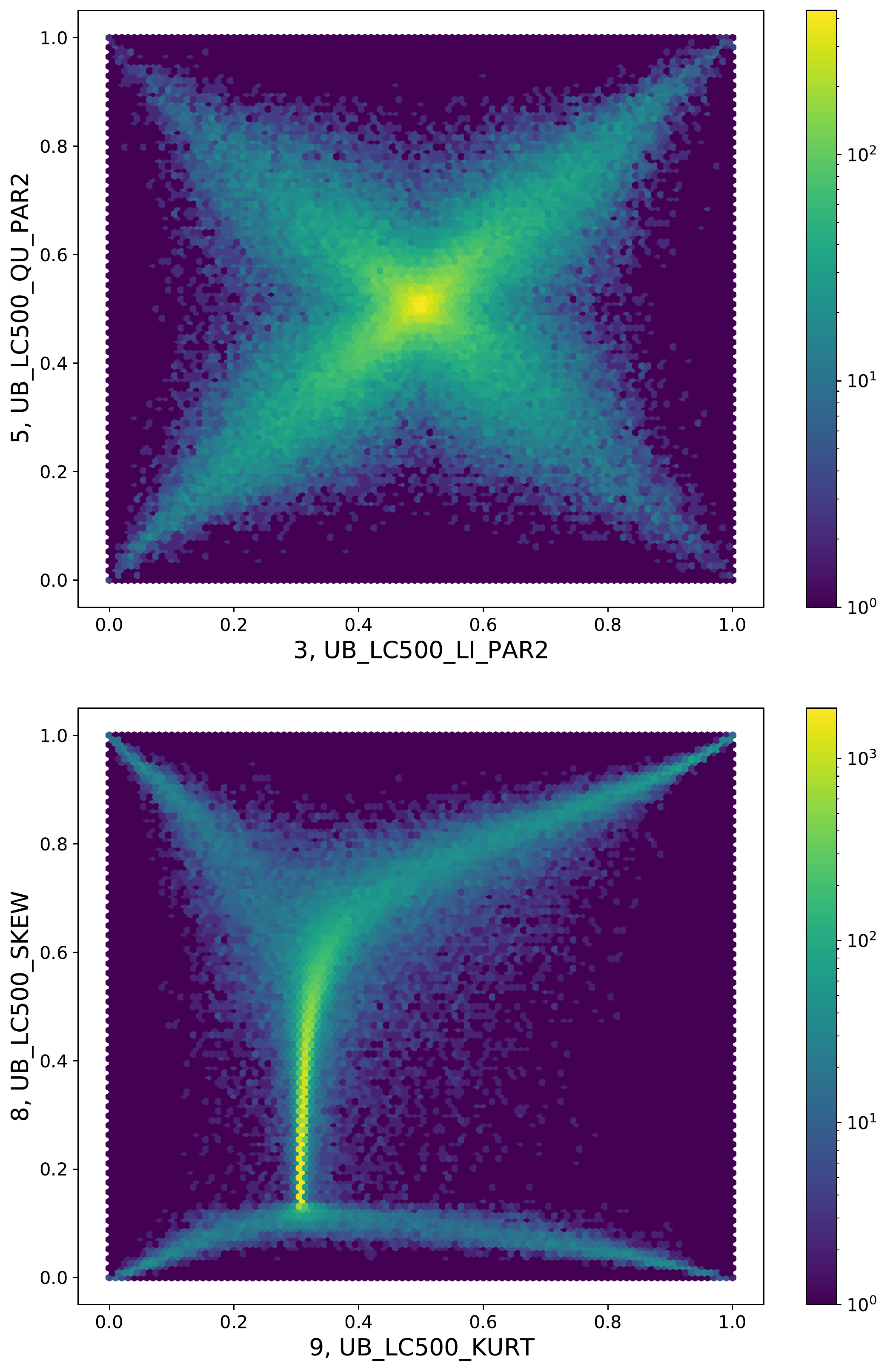}
    \caption{
    Two scatter plots illustrating nonlinear dependencies between parameters. As there are $n = 128\,925$ detections, the scatter plot is presented as a density plot. The values covered by the color bar, in logarithmic scale, present the number of detections in a given discrete area of the plot. Further explanation of the parameter correlations is given in the text.}
    \label{fig:sctr}
\end{figure}

The Pearson r correlation coefficient cannot accurately describe complicated nonlinear dependencies. Some of the more obvious examples are shown in Fig.~\ref{fig:sctr}. ~In the upper panel is a scatter plot of linear coefficient for linear and quadratic model (UB\_LC500\_LI\_PAR2 and UB\_LC500\_QU\_PAR2). Their correlation coefficient is almost zero, but there is a clear X-shaped dependence between these two parameters (the center corresponds to zero values of the original parameters). Two diagonal correlations have very similar absolute values but opposite signs, and cancel each other out producing a global coefficient close to zero. ~In the lower panel is a scatter plot between skewness and kurtosis. Their correlation coefficient is $\approx0.35,$ but there is a clear dependence, similarly to the previous case, but with several groups instead of a symmetric "X."\\

It is common to exclude redundant parameters in the machine learning process for several reasons, such as: the algorithm is more stable and faster with fewer parameters, and how each parameter affects the learning process is easier to interpret. 

The redundant parameters are typically those with high correlation to a given parameter. In this case, the correlation between parameters is highly complicated and it is not straightforward to exclude them based on a simple criterion. In some cases, high correlation is the result of very high positive correlation and a small negative correlation. If only one parameter were to be chosen, information would be lost from the negatively correlated part.
The SOM algorithm used in this work is relatively fast with this data set and there is no particular need to increase its speed efficiency by excluding parameters. 

Redundancy to a given parameter increases the dimensionality of the parameter space but does not contribute significantly to the information that the given parameter carries. As SOM is a dimensionality-reduction algorithm, it takes care of this naturally. The issue is that if there are more parameters in a group of correlated parameters, then the influence of the information from that group on the SOM learning process is increased. This is because the SOM "sees" the data in parameter space based on Euclidean distance. Therefore, this effect is approximately proportional to the square root of the number of redundant parameters, which is why it is not drastically important.
Based on all of the above, we decided to train the SOM with all the $m = 31$ normalized parameters.

\begin{table*}[htb]
\caption{Selected parameters.}
\label{tab:params}
\centering
\begin{tabular*}{0.95\textwidth}{l p{0.5\textwidth} c}
\hline\hline
Parameter designation & Parameter description & Narrowness\\
\hline
UB\_LC500\_CO\_PVAL\tablefootmark{a} & Tail probability for a constant model. & 3.77 \\
UB\_LC500\_LI\_PVAL\tablefootmark{a} & Tail probability for a linear model. & 3.71 \\
UB\_LC500\_LI\_PAR2 & Best-fit value of parameter 2 (the linear coefficient) for a linear model. & 85.5 \\
UB\_LC500\_QU\_PVAL\tablefootmark{a} & Tail probability for a quadratic model. & 3.70 \\
UB\_LC500\_QU\_PAR2 & Best-fit value of parameter 2 (the linear coefficient) for a quadratic model. & 58.9 \\
UB\_LC500\_QU\_PAR3 & Best-fit value of parameter 3 (the quadratic coefficient)
for a quadratic model. & 118\\
UB\_LC500\_STDEV & Weighted standard deviation on the distribution of the
rate. & 24.7 \\
UB\_LC500\_SKEW & Weighted skewness on the distribution of the rate. & $1.98 \times 10^7$ \\
UB\_LC500\_KURT & Weighted reduced kurtosis on the distribution of the rate. & $3.68 \times 10^{10}$ \\
UB\_LC500\_RELVAR & Relative variance (variance/average) on the distribution
of the rate. & $1.16 \times 10^3$ \\
UB\_LC500\_AMPLIT & Amplitude of rate excursion ((max(rate)-min(rate))/2). & 17.7 \\
UB\_LC500\_MEDABSDEV & Median absolute deviation of the distribution of the rate. & 24.6 \\
UB\_LC500\_MEDMAXOFF & Maximum relative offset from the median (max(|rate-
median|)/median) of the distribution of the rate. & 37.2 \\
UB\_CDF\_TFRAC\_BEL1S & Fraction of time spent more than 1 sigma below the
average rate. & 0.89 \\
UB\_CDF\_TFRAC\_ABO1S & Fraction of time spent more than 1 sigma above the
average rate. & 0.88 \\
UB\_CDF\_TFRAC\_BEL3S & Fraction of time spent more than 3 sigma below the
average rate. & 1.01 \\
UB\_CDF\_TFRAC\_ABO3S & Fraction of time spent more than 3 sigma above the
average rate. & 0.93 \\
UB\_CDF\_TFRAC\_BEL5S & Fraction of time spent more than 5 sigma below the
average rate. & 0.97 \\
UB\_CDF\_TFRAC\_ABO5S & Fraction of time spent more than 5 sigma above the
average rate. & 0.90 \\
UB\_CDF\_TFRAC\_MID20 & Fraction of time spent within 10 percent of the median
rate. & 1.62 \\
UB\_CDF\_RRANGE\_90 & Width of the range of rates in which the source spends 90
percent of its time. & 18.6 \\
UB\_CDF\_RFRAC\_MID20 & Fraction of UB\_CDF\_RRANGE\_90 in which the source
spends 20 percent of its time. & 1.27 \\
UB\_CDF\_RFRAC\_MID35 & ... 35 percent of its time. & 1.23 \\
UB\_CDF\_RFRAC\_MID50 & ... 50 percent of its time. & 1.04  \\
UB\_CDF\_RFRAC\_MID65 & ... 65 percent of its time. & 1.08 \\
UB\_CDF\_RFRAC\_MID80 & ... 80 percent of its time. & 1.06 \\
UB\_CDF\_RASYM\_MID20 & Asymmetry of the rate distribution in which the source
spends 20 percent of its time. & 46.0 \\
UB\_CDF\_RASYM\_MID35 & ... 35 percent of its time. & 41.6 \\
UB\_CDF\_RASYM\_MID50 & ... 50 percent of its time. & 70.4 \\
UB\_CDF\_RASYM\_MID65 & ... 65 percent of its time. & 73.7 \\
UB\_CDF\_RASYM\_MID80 & ... 80 percent of its time. & 72.5 \\
\hline
\end{tabular*}
\tablefoot{
Parameters used in training. All parameters were derived from light curves with 500 s uniform time bins. First column is a designation of the parameter in the WP2 catalog. Second column is the description of the parameter. Third column is the ratio of standard deviation to median absolute deviation.\\
\tablefoottext{a}{Tail probabilities (p-values) were recalculated with higher precision and transformed into one-sided sigma values such that higher sigma corresponds to a poorer model fit.}
}
\end{table*}

\section{Eclipsing sources from literature}
\label{app:res}

Table~\ref{tab:literature} contains eclipsing-like sources from the literature mentioned in Sect.~\ref{sec:dip}.

\begin{table*}[htb]
\caption{Eclipsing-like sources from the literature.}
\label{tab:literature}
\centering
\begin{tabular}{lcccc}
\hline\hline
Name & Reference & Obs. id & Src. num.\tablefootmark{a} & Pixel\\
\hline
RX J0047174-251811 & 1 & 0110900101 & 7 & 45,28\\
EP Dra & 2 & 0109464501 & 1 & 45,37\\
V* UY vol & 3 & 0560180701 & 1 & 45,32\\
 & & 0605560401 & 1 & 45,32\\
 & & 0651690101 & 1 & 45,32\\
 & & 0651690101 & 1 & 15,40\\
4U 1323-62 & 4 & 0109100201 & 1 & 45,40\\
V2301 Oph & 5 & 0109465301 & 1 & 45,38\\
4U 2129+47 & 6 & 0502460101 & 1 & 45,31\\
 & & 0502460201 & 1 & 45,31\\
 & & 0502460301 & 1 & 45,31\\
 & & 0502460401 & 1 & 45,31\\
2XMMp J131223.4+173659 & 7 & 0200000101 & 1 & 45,36\\
XTE J1710-281 & 8 & 0206990401 & 1 & 45,33\\
NGC 4736 ULX1 & 9 & 0094360601 & 1\tablefootmark{b} & 45,32\\
 & & 0094360601 & 2\tablefootmark{b} & 45,31\\
CAL 87 & 10 & 0153250101 & 1 & 45,33\\
AX J1745.6-2901 & 11 & 0402430301 & 5 & 45,33\\
 & & 0402430401 & 5 & 45,32\\
 & & 0402430701 & 5 & 45,32\\
 & & 0505670101 & 4 & 45,32\\
ULX CG X-1 & 12 & 0111240101 & 1 & 45,33\\
\hline
\end{tabular}
\tablefoot{
List of eclipsing-like sources randomly selected from the literature observed by {\it XMM-Newton}. We report the name, reference, observation and source number, and pixel coordinates in the BMU map (Fig.~\ref{fig:bmu_var}). Several detections may belong to the same source.\\
\tablefoottext{a}{Source number refers to 3XMM-DR4 notation.}
\tablefoottext{b}{Here the same source is detected as two different point-like sources in 3XMM-DR4.}
\tablebib{
(1) \citet{2003AA...402..457P}; (2) \citet{2004MNRAS.354..773R}; (3) \citet{2004MmSAI..75..484B}; (4) \citet{2005AA...436..195B}; (5) \citet{2007MNRAS.379.1209R}; (6) \citet{2007AA...476..301B}; (7) \citet{2008AA...485..787V}; (8) \citet{2009AA...502..905Y}; (9) \citet{2013ApJ...779..149L}; (10) \citet{2014ApJ...792...20R}; (11) \citet{2018MNRAS.477.3480J}; (12) \citet{2019ApJ...877...57Q}.
}
}
\end{table*}

\end{appendix}


\end{document}